\documentclass[11pt, a4paper]{article}
\usepackage{amsmath, amsfonts, amssymb}
\usepackage{graphicx,epsfig}
\usepackage{color}
\usepackage{hyperref}
\catcode`\=13
\def{$\bowtie$}
\begin{document}
\title{\textbf{Violation of the phase space general covariance as a diffeomorphism anomaly in quantum mechanics}}
\author{Nikolay Dedushenko\footnote{E-mail: dedushenko@gmail.com}\\\emph{Department of Physics, Taras Shevchenko University of Kiev,}\\\emph{ 2 Glushkova av. building 1, Kiev, Ukraine}\\
\emph{Department of Mathematics, Higher School of Economics,}\\
\emph{7 Vavilova Str., Moscow, Russia}} \maketitle \abstract{We
consider a topological quantum mechanics described by a phase space
path integral and study the 1-dimensional analog for the path
integral representation of the Kontsevich formula. We see that the
naive bosonic integral possesses divergences, that it is even
naively non-invariant and thus is ill-defined. We then consider a
super-extension of the theory which eliminates the divergences and
makes the theory naively invariant. This super-extension is
equivalent to the correct choice of measure and was discussed in the
literature. We then investigate the behavior of this extended theory
under diffeomorphisms of the extended phase space and despite of its
naive invariance find out that the theory possesses anomaly under
nonlinear diffeomorphisms. We localize the origin of the anomaly and
calculate the lowest nontrivial anomalous contribution.}
\newpage
\tableofcontents
\section{Introduction}
The notion of classical symmetry is often changed when proceeding to
the quantum world. Depending on the way one fixes the quantum
theory, the symmetry can be preserved, deformed or broken. In the
path integral approach the symmetries are treated in the most
classical-like way and the freedom of quantization consists in the
freedom of choice of regularization, while in an operator approach
this freedom consists in the choice of ordering (in both approaches
quantum corrections are possible as well).

When the symmetry can not be preserved one gets a quantum anomaly -
a phenomenon widely known in the context of quantum field theories.
Although this is usually connected with field theories in the
even-dimensional space-time, in the case of 1-dimensional QFT -
quantum mechanics - one still can find interesting effects that can
be treated as anomalies.

This is closely related to another possibility for the classical
symmetries in quantum world - possibility to be deformed. In such
case the classical symmetry seems to be broken and this can be
treated as the anomaly, however symmetry can be restored by adding
quantum corrections and thus replacing the notion of classical
symmetry by some quantum analog.

In the quantum mechanical case we know that the symmetries of the
classical Poisson manifold are deformed at the quantum level in
general. Namely, in the framework of deformation quantization one
may be interested in the symmetries of a particular star-product.
The classical symmetries form the subgroup inside classical
diffeomorphisms of the Poisson manifold and the classical
diffeomorphisms do not preserve the star-product. Instead they
change the star-product into the gauge-equivalent one (see the
corresponding paper by Kontsevich \cite{Ko1}).

The notion of diffeomorphism can be deformed via quantum corrections
in order to make the star-product covariant under such modified
transformations (see paper \cite{Lo1} where an appropriate
modification is described and discussed in terms of
$L_{\infty}$-morphism of Kontsevich). This may indicate that the
theory contains anomalies (as understood in the usual sense - the
theory is non-invariant under classical transformations).

We are going to investigate this in a current paper. The theory
under consideration is a topological quantum mechanics with a
classical limiting Hamiltonian system on a Poisson manifold
(Hamiltonian is taken to be 0 -- that is the meaning of
``topological''). We are going to analyze the covariance of this
system. Covariance of the system can be thought of as a special kind
of symmetry which means that the objects of the theory behave in a
good geometric way like in the classical case (e.g. like tensors).
If we had the covariance preserved then there would exist a
covariant formula for the star-product in terms of the Poisson
structure $\theta^{ij}(x)$ (i.e. the form of this formula in terms
of $\theta^{ij}(x)$ would be invariant). However such formula
doesn't exist, as we have stated before, which indicates the
breakdown of covariance.

We also may think in a following way. Kontsevich's deformation
quantization is formulated in terms of the 2-dimensional QFT with
quantum-mechanical observables living on the boundary (see
\cite{Cat} for the path integral formulation). However it would be
instructive to reformulate the quantization procedure in terms of
the 1-dimensional QFT, quantum mechanics by itself. And one could
expect that we have such an approach - a naively covariant path
integral. Indeed, we have the simplest star-product - a famous Moyal
product (which works for the case of the constant Poisson structure
$\theta^{ij}=const$):
\begin{equation}{
\left( f\star_M g\right)(x)=f(x) e^{{i\hbar\over
2}\overleftarrow{\partial_i}\theta^{ij}\overrightarrow{\partial_j}}
g(x) }
\end{equation}
Now consider a mechanical system with phase space $\mathbb{R}^d$
(where $d$ is even), with time taking values on a circle $t\in S^1,\
t\in[-\pi,\pi]$ and with the Poisson structure $\theta^{ij}=const$.
A well-known fact (see e.g. \cite{Cat} or \cite{Sza}) is that the
Moyal product can be written in terms of the path integral as a
special correlator of the topological quantum mechanics, namely as
follows:
\begin{equation}{
\label{moyal_product} \left( f\star_M g\right)(x)= {\int
\mathcal{D}\phi \prod_{i=1}^d\delta(\phi^i(\pm\pi)-x^i)
f\left(\phi(t_1)\right)g\left(\phi(t_2)\right) e^{{i\over\hbar}\int
{1\over 2}\omega_{ij}\phi^i\dot\phi^j dt} \over \int \mathcal{D}\phi
\prod_{i=1}^d\delta(\phi^i(\pm\pi)-x^i) e^{{i\over\hbar}\int {1\over
2}\omega_{ij}\phi^i\dot\phi^j dt} } }
\end{equation}
where $\omega$ is the symplectic structure, i.e. it is inverse to
the Poisson structure: $\omega_{ij}\theta^{jk}=\delta_i^k$. But this
gives us a way to make a straightforward but naive generalization of
the Moyal product to the case of the non-constant\footnote{In fact
one can provide a naive reasoning for this formula. In \cite{Cat}
the following path integral representation for the Kontsevich
quantization formula had been obtained: $$ \big(f\star g\big) (x) =
\int_{X(\infty)=x} f(X(1))g(X(0))e^{{i\over\hbar}S[X,p]}
$$
with fields $X$ and $p$ living on a disk $D^2$, three points $0, 1,
\infty$ fixed on a boundary of the disk and with action
$$
S[X,p]=\int_{D^2} p_i\wedge dX^i + {1\over 2} \theta^{ij} p_i\wedge
p_j
$$
If one formally integrates over $p$'s in this formula then all the
bulk-dependence naively drops out and we are left with a boundary
theory:
$$ \big(f\star g\big) (x) = \int_{X(\infty)=x}\widetilde{\mathcal{D}X}
f(X(1))g(X(0))e^{{i\over\hbar}\int \alpha}
$$
with $d\alpha=\omega$ and with a special naive measure
$\widetilde{\mathcal{D}X} = \prod_{t\in\partial D^2}
\sqrt{\det\omega\big(X(t)\big)}d^dX(t)$. Although this is not
exactly the formula (\ref{naive_product}), but rather the formula to
be discussed in the ``Super improvement'' section, it is instructive
to start with a ``wrong'' formula (\ref{naive_product}) to see why
it fails and only then study the naively correct one.
 } $\theta^{ij}$:
\begin{equation}{
\label{naive_product} \left( f\star g\right)(x)= {\int
\mathcal{D}\phi \prod_{i=1}^d\delta(\phi^i(\pm\pi)-x^i)
f\left(\phi(t_1)\right)g\left(\phi(t_2)\right) e^{{i\over\hbar}\int
\alpha} \over \int \mathcal{D}\phi
\prod_{i=1}^d\delta(\phi^i(\pm\pi)-x^i) e^{{i\over\hbar}\int \alpha}
}}
\end{equation}
where $\alpha$ is a 1-form such that $d\alpha=\omega$. From the
naive point of view this formula could give us a covariant way to
define the star-product for the case of an arbitrary
$\theta^{ij}(x)$. And naively it could be invariant under
diffeomorphisms. But we have discussed that there is no way to write
the star-product on the symplectic (or Poisson) manifold covariant
under \emph{classical} diffeomorphisms without introduction of some
additional structures.

This indicates again that ($\ref{naive_product}$) is non-invariant
in fact and contains some kind of anomaly responsible for this
non-invariance.

The goal of the current paper is to built the corresponding
framework and to study the anomaly in the simplest case where it
shows itself. We are not going to deal with the whole expression
($\ref{naive_product}$) yet, neither to check if it really is a
star-product (i.e. if it is associative) - that will be the topic of
the upcoming research.

In the chapter ``Path integral for the phase space'' it will be
shown that the object ($\ref{naive_product}$) is an ill-defined
object in fact. A supersymmetric improvement of it will be provided
as well as the regularization which is always crucial for the path
integral. After such improvements ($\ref{naive_product}$) will
become well-defined and finite.

We'll provide some illustrative calculations in the chapter
``Examples'', namely we'll prove the formula
($\ref{moyal_product}$), check the classical limit of
($\ref{naive_product}$) and show peculiarities of loop calculations
in our theory.

The chapter ``Anomaly chasing'' is devoted to the understanding of
how the diffeomorphism acts in our theory. We'll discuss how to
perform diffeomorpism compatible with the regularization describing
the two approaches for this - the naive one and the proper one.
We'll obtain the anomaly in the lowest order of perturbation theory
in the naive
approach and then we'll explain its nature in the proper approach.\\
\\
\textbf{Acknowledgements.} The author is grateful to Andrei Losev
for advices and discussions that played a crucial role in writing of
this paper. The author is also grateful to Oleksandr Gamayun for
useful remarks.
\section{Path integral for the phase space}
\subsection{Naive bosonic approach}
\subsubsection{Formulation}
In order to build a framework we should define a path integral
$\int\mathcal{D}\phi e^{{i\over\hbar}\int\alpha}$ at first, where
$\phi^i(t)$ is a map $\phi: S^1 \to M$ where $M$ is the
$d$-dimensional symplectic manifold ($d$ is even) with symplectic
form $\omega=d\alpha$ and $S^1$ is the time manifold.

We'll provide a standard perturbation theory later and thus
decompose the action into the quadratic and non-quadratic parts. But
such decomposition makes sense only when $M=\mathbb{R}^d$ - the
property of being quadratic is purely coordinate and has no global
meaning in the case of an arbitrary manifold. So we need to consider
$M=\mathbb{R}^d$ as soon as we need a perturbation theory. There's
also another reason for such a restriction related with measure.

We have to define measure $\mathcal{D}\phi$ in some way. There are
basically two ways to do that - we can either provide time-slicing
(i.e. lattice regularization) or provide mode decomposition. We are
not going into details here - we just mention that the first
procedure is ill-defined and needs some extra analysis in the case
of the phase space. So we choose the second one according to which
we have to provide Fourier decomposition of the fields:
\begin{equation}{
\label{fourier1} \phi^i(t)=\sum_{n=-\infty}^{\infty}\phi^i_n e^{i n
t} }
\end{equation}
where $\phi^i_n \in \mathbb{C}$ and $\phi^i_{-n} =
\overline{\phi^i_n}$ (the requirement of reality of $\phi^i(t)$) and
then construct the measure in an appropriate way:
\begin{equation}{
\label{naive_measure} \mathcal{D}\phi \sim \prod_{n} \prod_i
d\phi^i_n = \prod_i d\phi^i_0\prod_{n>0} \prod_i d\phi^i_n
d\overline{\phi^i_n}}
\end{equation}
regardless if it is well-defined yet.

Possibility to make this is the second evidence for the manifold $M$
to be just $\mathbb{R}^d$. Indeed, Fourier decomposition
($\ref{fourier1}$) makes no sense until we can take linear
combinations of the fields, which means that $M$ should be a linear
space.

Next we put:
\begin{equation}{
\label{alpha_form} \alpha = {1\over 2}\omega^{(0)}_{ij}\phi^i
d\phi^j + e_i(\phi)d\phi^i }
\end{equation}
\begin{equation}{
\label{symplectic_form} \omega=d\alpha =
\omega^{(0)}_{ij}d\phi^i\wedge d\phi^j + de }
\end{equation}
where $\omega^{(0)}_{ij}$ is constant and provide perturbation
theory expansions in $e_i(\phi)$.
\subsubsection{Non-invariance of the measure}
At this point we can already notice that measure
($\ref{naive_measure}$) is by no means invariant under
diffeomorphisms of the symplectic manifold. Indeed, if we provide a
coordinate change, the Jacobian will appear and the general
covariance will be broken: $\mathcal{D}\phi \sim \left|{\delta
\phi\over \delta\phi'}\right|\mathcal{D}\phi'$ where $\left|{\delta
\phi\over \delta\phi'}\right|$ is a non-constant functional and thus
it contributes into the correlators. So the answer obtained within
such a prescription is not expected to be covariant.
\subsubsection{Study of ultraviolet divergences}
However, non-invariance is not the only pathology of the described
object - the case is that it is divergent as we'll see later and
even non-renormalizable in a standard QFT sense.
\\
Indeed, introduce diagram notations for the propagators of the
non-interacting theory (arrow notation for the derivative is similar
to the fat dot notation from the book \cite{BasNie}):
$$
^0\left\langle\left(\phi^i(t_1)-x^i\right)\left(\phi^j(t_2)-x^j\right)\right\rangle
= \epsfig{file=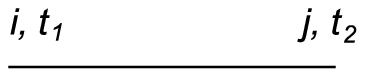}
$$
$$
^0\left\langle\left(\phi^i(t_1)-x^i\right)\dot\phi^j(t_2)\right\rangle
= \epsfig{file=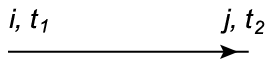}
$$
$$
^0\left\langle\dot\phi^i(t_1)\dot\phi^j(t_2)\right\rangle =
\epsfig{file=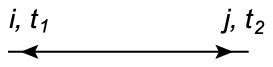}
$$
where notation $^0\langle...\rangle$ means that correlator under
consideration is the free one, i.e. of the quadratic theory. As soon
as the action is of the first-order type $S_0=\int dt {1\over
2}\omega_{ij}\phi^i\dot\phi^j$, we can conclude that in the momentum
space the first propagator is proportional to the inverse momentum
${1\over p}$, the second one is $\sim 1$ and the third one is $\sim
p$. This means that every wheel-like diagram with equal number of
internal derivatives (denoted by arrows) and internal propagators:
$$
\epsfig{file=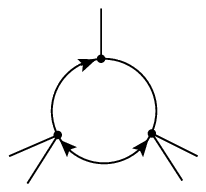}
$$
is linearly divergent. Diagrams of the type:
$$
\epsfig{file=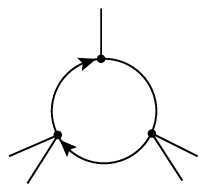}
$$
where the number of internal propagators exceeds the number of
internal arrows by $1$, are logarithmical divergent, but here we can
notice that in the case of an odd-dimensional world-sheet (which is
our case - it is 1-dimensional in our theory) logarithmical
divergences do not occur due to the symmetries of the integrand.
Indeed, expression of the type $\int_{a<|p|<\Lambda}{d^n p\over
p^n}$ is $\propto \log\Lambda$ for even $n$ and equals to 0 for odd
$n$. In the case of odd $n$ the result depends on the form of the
cut-off and thus the divergence is replaced by the ambiguity. But
existence of an infinite series of the linear divergent wheel-like
diagrams indicates that the theory is not just divergent but even
non-renormalizable. One can find more detailed discussion in the
Appendix A.

So what do we see? Theory that looks very natural at the first sight
happened to be extremely ill-defined. Although it was 1-dimensional
it happened to be non-renormalizable due to the interactions
containing derivatives. Fortunately all the troubles can be solved
in a one simple step - that is the topic of the following
subsection.
\subsection{Super improvement}
We can restore measure invariance and theory finiteness by
introducing a generally-covariant measure:
\begin{equation}
{ \label{covar_measure} \mathcal{D}\phi \sim \prod_t
\sqrt{\det\omega\left(\phi(t)\right)}d\phi^1(t)...d\phi^d(t) }
\end{equation}
where $d$ is even. A piece of luck is that all the needed
counterterms to cancel the divergences are present in this choice of
measure. To see that we'll lift the square root of $\det\omega$ by
introducing an anticommuting real \emph{ghost} field $\psi^i$ (in
the spirit of \cite{Bast} and \cite{BasNie} where the ghosts are
intensively used in the configuration space theory):
\begin{equation}{
\label{ghosts} \sqrt{\det\omega\left(\phi(t)\right)}\sim \int
d\psi^1(t)...d\psi^d(t) e^{{i\over\hbar}
\omega_{ij}\left(\phi(t)\right)\psi^i(t)\psi^j(t)} }
\end{equation}
and thus consider a modified super measure
$\mathcal{D}\phi\mathcal{D}\psi$.

Such measure has already been talked about in the literature in
different contexts. Connection of covariant measure
(\ref{covar_measure}) and supersymmetric measure
$\mathcal{D}\phi\mathcal{D}\psi$ was discussed in (\cite{Nico1}) and
(\cite{Nico2}) and essentially used in (\cite{Morozov}).

One can notice without further considerations that an obtained
extended theory with an action $S=\int dt\
\big\{\alpha_i(\phi)\dot\phi^i +
\omega_{ij}(\phi)\psi^i\psi^j\big\}$ possesses supersymmetry, i.e.
it is invariant under:
$$
\delta \phi^i = \theta \psi^i,\qquad \delta\psi^i = -\theta
\dot\phi^i
$$
with $\theta$ being small anticommuting parameter (see
(\cite{Morozov}) again). However, we'll not need this property in a
further consideration at all.

The main advantages of the super improvement are manifest measure
invariance and divergences cancelation. In this subsection we are
going to define a quantum mechanical path integral based on such a
modified approach and check its consistency.
\subsubsection{The construction}
Suppose we have a symplectic manifold $(M, \omega)$ where $M$ is a
d-dimensional (d is even) smooth manifold and $\omega$ is a
symplectic form. We have already discussed before that there should
be a linear structure on $M$ in order to define a perturbative path
integral with mode-regularized measure, so we consider
$M=\mathbb{R}^d$. We also have a time manifold $S^1$. Let us
parameterize it by $t\in [-\pi,\pi]$. Consider a continuous and thus
integrable map:
\begin{equation}{
\label{super_map} \phi: S^1 \to \mathbb{R}^d }
\end{equation}
This map is just a $d$-component real boson field on $S^1$. We
describe it in terms of its Fourier transform:
\begin{equation}{
\label{fourier_bosonic} \phi^k(t)=\sum_{n\in \mathbb{Z}} \phi^k_n
e^{i n t}\ \ \ \ (k=1..d) }
\end{equation}
where $\phi^i_n$ are complex bosons with
\begin{equation}{
\label{reality_bos} \phi^i_{-n}=\overline{\phi^i_n} }
\end{equation}
We also consider a $d$-component fermion field $\psi^i(t),\ i=1..d$.
We define it as a set of $d$ time-dependent linear combinations in
the infinite-dimensional Grassmann algebra:
\begin{equation}{
\label{fourier_fermionic} \psi^k(t)=\sum_{n\in \mathbb{Z}} \psi^k_n
e^{i n t}\ \ \ \ (k=1..d) }
\end{equation}
where $\psi^i_n$ are fermion Fourier modes - the complex Grassmann
variables (except of $\psi^i_0$ which is real) satisfying:
\begin{equation}{
\label{reality_ferm} \psi^i_{-n}=\overline{\psi^i_n}. }
\end{equation}
Here we should notice that due to the reality condition
($\ref{reality_bos}$) the zeroth boson mode $\phi^i_0$ is real while
$(\phi^i_1, \phi^i_{-1}=\overline{\phi^i_1})$, $(\phi^i_2,
\phi^i_{-2}=\overline{\phi^i_2})$, etc. each takes values in
$\mathbb{C}$. But as integration over fermions has a formal
algebraic meaning (as opposed to the boson integrals that have
analytical sense) we can think of them due to ($\ref{reality_ferm}$)
in two equivalent ways: either say that $\psi^i_0$ is real and
$(\psi^i_1,\psi^i_{-1}=\overline{\psi^i_1})$,
$(\psi^i_2,\psi^i_{-2}=\overline{\psi^i_2})$ etc. are complex, i.e.
each form $\Pi\mathbb{C}$ or say that all the fermions $...,\
\psi^i_{-2},\ \psi^i_{-1},\ \psi^i_0,\ \psi^i_1,\ \psi^i_2,...$ are
real. We will prefer the first understanding due to its analogy to
the bosonic case.

We define the mode-regularized fields as follows:
$$
\phi^i_N(t) = \sum_{n=-N}^N\phi^i_n e^{i n t}
$$
\begin{equation}{
\label{fourier_reg} \psi^i_N(t) = \sum_{n=-N}^N\psi^i_n e^{i n t} }
\end{equation}
Notice that regularized bosons take values in $\mathbb{R}^d\times\mathbb{C}^{Nd}$, where $\phi^i_0\in\mathbb{R}$ and $(\phi^i_k,\phi^i_{-k})\in\mathbb{C}$.\\
Fix the action functional:
\begin{equation}{
\label{action_main} S[\phi,\psi] = \int\left\{
\alpha_i\left(\phi\right)\dot\phi^i +
\omega_{ij}\left(\phi\right)\psi^i\psi^j - H\left(\phi\right)
\right\}dt }
\end{equation}
where $\omega=d\alpha$ and take a functional $F[\phi,\psi]$.\\
\textbf{Definition 1.} Mode-regularized functional integral in the
phase space is an expression:
\begin{equation}{
\label{pathint_definition} I_N[F,S] = \mathcal{A}(N)\int
\prod_{n=-N}^N\left[d\phi_n^1...d\phi_n^d
d\psi_n^1...d\psi_n^d\right]
F[\phi_N,\psi_N]e^{{i\over\hbar}S[\phi_N,\psi_N]} }
\end{equation}
where integration over $\psi$'s is understood in the sense of
Berezin, integration over $\phi$'s is provided throughout all the
space $\mathbb{R}^d\times\mathbb{C}^{Nd}$ and $\mathcal{A}(N)$ does
not depend on $F[\phi,\psi]$.

We'll say that functional integral exists in the sense of
mode-regularization if one can find such a function $\mathcal{A}(N)$
(that does not depend on $F[\phi,\psi]$) that $\lim_{N\to\infty}I_N$
exists.

In practice we will understand ($\ref{pathint_definition}$) in a perturbative way, namely we'll expand the non-quadratic part of the action into the Taylor series, integrate each term and understand existence of the path integral as existence of all the terms in the resulting formal series.\\
When we choose $\mathcal{A}(N)={1\over I_N[1,S]}$ we get absolute correlators (averaged quantities):\\
\textbf{Definition 2.}\\
1) An absolute regularized correlator of $F[\phi,\psi]$ is the
following:
\begin{equation}{
\label{abs_corr} {I_N[F,S]\over I_N[1,S]} }
\end{equation}
2) A relative regularized correlator of $F[\phi,\psi]$ for a given
functional $G[\phi,\psi]$ is the following:
\begin{equation}{
\label{rel_corr} {I_N[F G,S]\over I_N[G,S]} }
\end{equation}

The appropriate limits $N\to\infty$ (if they do exist) are called
absolute and relative correlators. Notice that if the absolute
correlator exists so does the relative one.

In our main case of interest the Hamiltonian will be equal to zero
$H(\phi)=0$ and we'll consider the following relative correlators:
\begin{equation}{
\label{corr_reg_before_N} \langle F[\phi,\psi] \rangle_N={I_N[F
\eta_x[\phi],S]\over I_N[\eta_x[\phi],S]}}
\end{equation}
\begin{equation}{
\label{corr_reg} \langle F[\phi,\psi] \rangle=\lim_{N\to\infty}
{I_N[F \eta_x[\phi],S]\over I_N[\eta_x[\phi],S]}}
\end{equation}
where $\eta_x[\phi]=\prod_i\delta\left(\phi^i(\pi)-x^i\right)$ and
we have introduced a widespread notation $\langle...\rangle$ for the
correlator.

Notice that in the classical limit $\eta_x[\phi]$ is nothing but an
"evaluation observable" - it evaluates the value of the inserted
observable $F[\phi]$ at the classical solution with
$\phi^i(\pi)=x^i$ (if such solution exists) - in our case such
solution really exists and it is just $\phi^i(t)=x^i=const$. If the
inserted observable is the product of functions: $F[\phi]=
f_1\left(\phi(t_1)\right)f_2\left(\phi(t_2)\right)...f_n\left(\phi(t_n)\right)$
then in the limit $\hbar\to 0$ ($\ref{corr_reg}$) gives rise to an
ordinary point-wise product of functions evaluated at the point $x$:
$f_1(x)f_2(x)...f_n(x)$. This will be proved in the section 3.2
where we'll provide the tree level calculation of such correlator.

From technical point of view $\eta_x[\phi]$ is an infrared regulator
(zero mode regulator) - we need it due to the action $\int_{S^1}
{1\over 2}\omega^{(0)}_{ij}\phi^i\dot\phi^j dt$ being invariant
under global translations $\phi^i \to \phi^i + c^i$.

Then we extract the quadratic part of the action in order to provide
perturbations. To do that we use
($\ref{alpha_form}$)-($\ref{symplectic_form}$) and get $S=S_0 +
S_{int}$ where:
\begin{equation}{
\label{background_alpha} S_0 = \int \left\{{1\over
2}\omega^{(0)}_{ij}\phi^i\dot\phi^j +
\omega^{(0)}_{ij}\psi^i\psi^j\right\} dt }
\end{equation}
\begin{equation}{
\label{background_omega} S_{int} = \int \left\{e_i(\phi)\dot\phi^i +
de_{ij}\psi^i\psi^j\right\} dt }
\end{equation}

Notice that the choose of background ($\ref{background_alpha}$) is not coordinate-invariant, however it is preserved by the linear transformations as well as the linear structure on $M$.\\
\textbf{Free correlators}

Consider free correlators, i.e. correlators of the theory with
action $S_0$. Introduce appropriate notations
$^0\langle...\rangle_N,\ ^0\langle...\rangle$ (such notation have
already been used in 2.1.3). We calculate them explicitly:
\begin{equation}{
\label{prop_bos} ^0\langle\phi_N^i(t_1)\phi_N^j(t_2)\rangle_{N} =x^i
x^j + i\hbar\theta^{ij}_{(0)} G_N(t_1,t_2) }
\end{equation}
Where $\theta_{(0)}=\left(\omega^{(0)}\right)^{-1}$,
$G_N(t_1,t_2)=\sum_{n,m = -N}^N G_{n,m} e^{i n t_1 + i m t_2}$ and:
$$
G_{n,m}={1\over 2\pi i}\left\{ {\delta_{n+m}\over n} -
{\delta_m\over n}(-1)^n + {\delta_n\over m}(-1)^m\right\}
$$
\begin{equation}{
\label{propagator} G_{0,0}=0 }
\end{equation}
The answer for $\psi$'s can be calculated in the same way and is
given by:
\begin{equation}{
\label{prop_ferm_mode} ^0\langle\psi^i_n \psi^j_n\rangle_{N} =
{i\hbar\over 2}\theta_{(0)}^{ij}{\delta_{n+m}\over 2\pi} }
\end{equation}
or equivalently:
\begin{equation}{
\label{prop_ferm_mode_equiv} ^0\langle\psi_N^i(t_1)
\psi_N^j(t_2)\rangle_{N} = {i\hbar\over
2}\theta_{(0)}^{ij}\delta_N(t_1-t_2) }
\end{equation}
where $\delta_N(t)$ is a regularized delta-function:
\begin{equation}{
\label{regular_delta} \delta_N(t) = {1\over 2\pi}\sum_{n=-N }^Ne^{i
n t} }
\end{equation}
In continuous limit:
\begin{equation}{
\label{prop_ferm_cont}
^0\langle\psi^i(t_1)\psi^j(t_2)\rangle={i\hbar\over
2}\theta_{(0)}^{ij}\delta(t_1-t_2) }
\end{equation}
And in ($\ref{prop_bos}$):
\begin{equation}{
\label{propagator_cont} G(t_1,t_2)=\begin{cases}{1\over 2}Sign(t_1 -
t_2),&if\ -\pi<t_1,t_2<\pi;\cr 0,&if\ t_1=\pi\ {\rm or}\
t_2=\pi.\cr\end{cases} }
\end{equation}
We should also notice that $G_N(t_1,t_2)=-G_N(t_2,t_1)$ and:
\begin{equation}{
\label{derivative_of_prop} {\partial G_N(t_1,t_2)\over \partial t_1
}=\delta_N(t_1-t_2) - \delta_N(t_1-\pi) }
\end{equation}
Now we can use perturbation theory in order to compute correlation
functions up to the fixed order in couplings.

It is important to note that when calculating perturbations we take
self-contractions into account. Indeed, from one point of view there
is no reason to neglect them. From another point of view if we
ignored such terms we could possibly break covariance under
diffeomorphisms and thus it would need additional analysis which we
don't want to provide for the purposes of simplicity.\\ We'll use
the Feynman diagrams technics described above for convenience.
\subsubsection{Naive expectations}
The first pathology that is claimed to be got rid of by the
supersymmetric improvement is measure non-invariance. Let us observe
how it works from the naive point of view. From the expression
($\ref{action_main}$) we see that fermions should transform like
vector fields under diffeomorphisms (in order to make action a
scalar), namely if we provide:
\begin{equation}{
\label{big_diffeo_bos} \phi^i = \phi^i(\varphi) }
\end{equation}
then we should make a substitution in a fermion sector (here
$\widetilde{\psi}$ is a fermion):
\begin{equation}{
\label{big_diffeo_ferm} \psi^i = \widetilde{\psi}^j
{\partial\phi^i\over\partial\varphi^j} }
\end{equation}
But now we can observe that
\begin{equation}{
\label{naive_trans} d^d\phi(t) d^d\psi(t) =
\left[\det{\partial\phi^i(\varphi(t))\over\partial\varphi^j(t)}\right]
d^d\varphi
\left[\det{\partial\phi^i(\varphi(t))\over\partial\varphi^j(t)}\right]^{-1}d^d\widetilde{\psi}
= d^d\varphi d^d\widetilde{\psi}
 }
\end{equation}
and thus naively conclude that in the functional case product of the
standard boson measure and the Berezin fermion measure
$\mathcal{D}\phi\mathcal{D}\psi$``=''$\prod_t d^d\phi(t) d^d\psi(t)$
is invariant under diffeomorphisms. To get rid of the word
``naively'' one should reexamine it in a regularization (as the path
integral is meaningless without regularization). We'll return to
this question later when discussing the diffeomorphisms and
regularization - we will find out that the measure is really
invariant with some additional assumptions about diffeomorphism (to
be discussed in the Anomaly chasing chapter).

We've seen that the starting bosonic theory was divergent. Now we
claim that the supersymmetric improvement solves this problem - the
theory becomes finite.\\
We can provide a loose argument of finiteness here. Since the theory
is naively invariant under diffeomorphisms, we can find such
coordinates $p_i, q^i$ (as it is stated by the Darboux theorem),
that $\omega=\sum_{i=1}^{d/2}dp_i\wedge dq^i$ and thus we can choose
$\alpha=\sum_{i=1}^{d/2}p_i dq^i$ to make the action quadratic. But
quadratic theory is pretty well-defined - we don't need any
perturbations in it. So does the starting theory.

The only possible divergences can arise if we consider some special
divergent observables, e.g. $^0\langle\phi^i(t)\dot\phi^j(t)\rangle$
is divergent even in the quadratic theory. To get rid of such cases
we'll consider only observables of the type $f_1(\phi(t_1))
f_2(\phi(t_2))...f_n(\phi(t_n))$ further.

Our naive argument is straightforward but wrong as long as we don't
know weather the mode-regularized theory is still invariant with
respect to diffeomorphisms. Moreover, it will be shown that the
invariance is broken but it's breakdown is finite and in some sense
the argument above is not as bad.  However to provide a successive
theory we should find a rigorous argumentation at this point. We
will provide a detailed perturbative analysis to prove the
finiteness.

\subsubsection{Cancelation of divergences: loop analysis}
Consider correlation function of the monomial:
\begin{equation}{
\label{poly_corr} \langle\phi^{i_1}(t_1)...\phi^{i_k}(t_k)\rangle }
\end{equation}
We'll prove that ($\ref{poly_corr}$) is finite and piecewise-continuous (i.e. it can contain jumps but not infinities).\\
In order to do that we should describe diagram technics. It is
almost the same as before except of existence of anticommuting
ghosts. So we introduce notation for the free ghost propagator:
$$
^0\langle\psi^i(t_1)\psi^j(t_2)\rangle=\epsfig{file=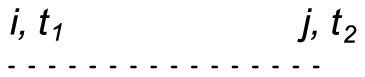}
$$
And introduce a new type of vertexes corresponding to the couplings
of $\psi$'s with $\phi$'s:
$$
\begin{matrix}\ \cr{i\over\hbar}\int dt\ \omega_{ij}(\phi)\psi^i\psi^j
=\cr\ \cr\end{matrix}\begin{matrix}\ \cr
\epsfig{file=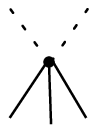}\cr\end{matrix}
$$
where the number of external bosons is not fixed as in general case
$e_i(\phi)\dot\phi^i$ interaction contains all the possible terms of
the type ${1\over
k!}\partial_{j_1}...\partial_{j_k}e_i(0)\phi^{j_1}...\phi^{j_k}\dot\phi^i$.
Now let us make several observations.\\
\textbf{Observation 1:} Ghosts can occur only in loops. This follows
immediately from that ghosts contribute quadratically to the
appropriate vertexes. Thus we can neglect ghosts on a tree level.
Moreover, every ghost loop is a circle with exterior bosonic legs.\\
\textbf{Observation 2:} Propagator of bosons ($\ref{propagator}$)
doesn't preserve momentum while vertexes as long as the propagator
of ghosts ($\ref{prop_ferm_mode}$) preserve it. This is natural
because we've broken translation invariance on a world line when
regularizing 0-mode by
the $\delta$-function.\\
\textbf{Observation 3:} All tree-level diagrams are finite. This
statement holds because the tree level reproduces the classical
answer which is definitely finite - the argumentation is standard
here. At the tree level one can safely take a limit $N\to\infty$ and
work with integrals of distributions - everything happens to be
well-defined then.\\
So we have to check possible loop diagrams and prove that their sum is finite.\\
\textbf{Observation 4:} There are no logarithmic divergences in
theories with an odd-dimensional world-sheet - instead we have
logarithmic ambiguities. We have already mentioned this in the
section 2.1.3.\\
So now all we need to do is to show that loops do not give rise to
the linear (or higher) divergences. There exist three types of loops
in our theory: ghost-loops, non-ghost-loops and loops containing
both ghost and non-ghost fields. But as soon as ghosts always form
loops, it is enough to show cancelation of such loops - then we'll
get rid of ghosts at all. Finiteness of correlators will follow from
this immediately.\\
A step-by-step loop analysis can be found in the Appendix B. Here we
only state that the main effect is that both divergent bosonic loops
and ghost loops contain $\delta_N(0)$ divergences, that are in fact
linear divergences as $\delta_N(0) \sim \sum_{-N}^N 1$. Then we find
that to every ghost loop there corresponds a set of divergent
bosonic loops with the same number of external legs. We evaluate
them for the finite $N$ and find that divergent parts (i.e.
expressions proportional to $N$) cancel out and we finally get what
we needed. We can take a limit $N\to\infty$ safely at the end. The
diagram illustration is as follows:
$$
\epsfig{file=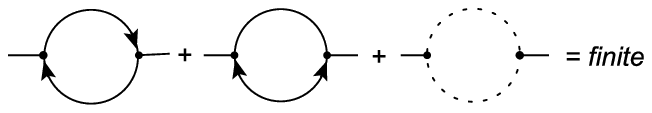}
$$
So we have shown that our super-improved approach gives rise to the
finite theory. However in the Appendix B this is discussed only in
the case of some special system - quantum mechanics on a circle with
zero hamiltonian and with delta-function fixing fields at $t=\pi$.
This can be generalized to an arbitrary non-zero hamiltonian in a
straightforward way. We'll not show this explicitly but only give an
idea.

Divergencies will always arise from the non-ghost loops with equal
number of internal propagators and derivatives (internal arrows).
The point is that propagator of the type
$\langle\phi^i(t)\phi^j(0)\rangle$ always contains jump at $t=0$ as
soon as $\lim_{t\to
+0}\left[\langle\phi^i(t)\phi^j(0)\rangle-\langle\phi^j(t)\phi^i(0)\rangle\right]=i\hbar\langle\theta^{ij}\rangle+O(\hbar^2)$
due to the usual quantum mechanical correspondence principle. This
jump gives rise to the $\delta$-function after differentiation which
causes the $\delta_N(0)$-divergency after all. And such divergencies
are exactly canceled by the ghost loops. Finally we get a finite
one-dimensional
quantum field theory as expected.\\
Notice that from the "Naive expectations" subsection we know that
finiteness follows from the invariance. We have shown that the
theory is finite. Then it still can be invariant or non-invariant.
We'll check which case is true in the "Anomaly chasing" section.
\section{Examples}
\subsection{Free correlation functions}
\subsubsection{Propagators}
The free propagators are described by the formulas
($\ref{prop_bos}$)-($\ref{propagator_cont}$). Notice that $G_{n,m}$
has such a form that
\begin{equation}{
\label{same_time} \sum_{n,m=-N}^N G_{n,m}e^{i(n+m)t}=0 }
\end{equation}
and thus in our prescription we have $G(t,t)=G_N(t,t)=0$. As we have
already mentioned, $G(t,\pi)=G(\pi,t)=0$ and so the propagator has
the structure as follows:
$$
\epsfig{file=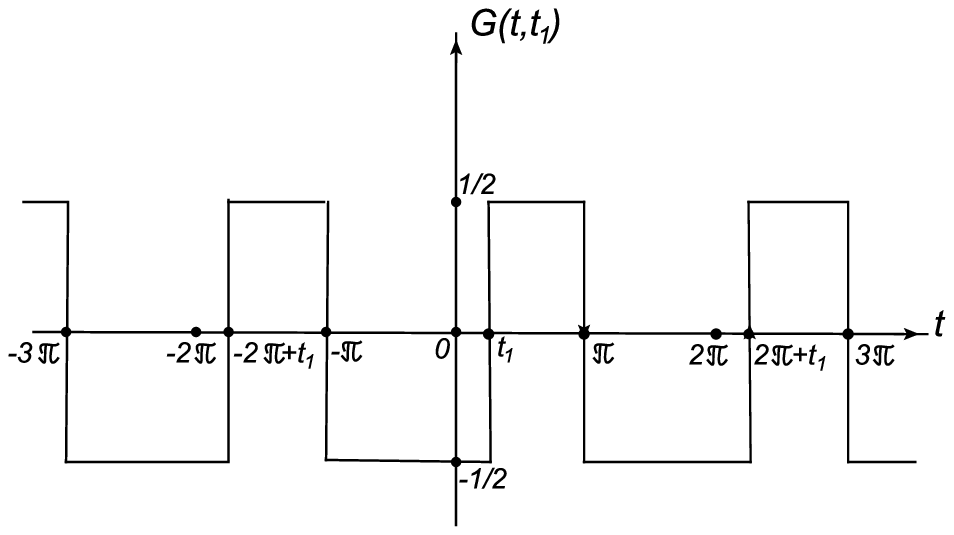}
$$
As this function is not smooth and we often deal with operations
that demand smoothness (or at least differentiability), we should
always be careful and work with regularized expression (with cut-off
parameter $N$) which is smooth and take $N\to\infty$ limit only at
the end.
\subsubsection{Regularized delta-function}
In a section 2 we have introduced a regularized delta-function
$\delta_N(t)={1\over 2\pi}\sum_{n=-N}^N e^{i n t}$. In the current
subsection we are going to collect some of its elementary properties
as long as they will be widely used below.

1) According to ($\ref{regular_delta}$) the function $\delta_N(t)$
is normalized in such a way, that
\begin{equation}{
\int \delta_N(t) dt = 1 }
\end{equation}

2) $\delta_N(t)$ acts like a real delta-function on the space of
functions, which Fourier transforms have the support lying inside
the set $\{-N,-N+1,...,N-1,N\}$, i.e. it acts like a real
delta-function on the space of Fourier polynomials $span\{e^{- i N
t}, e^{-i(N-1)t},...,e^{i N t}\}$. Indeed, if we have $P(t) =
\sum_{k=-N}^N p_k e^{i k t}$, then:
\begin{eqnarray}
\int P(t) \delta_N(t-t_0) dt &= \sum_{n,k=-N}^N p_k {1\over 2\pi}
\int e^{i(k+n)t-i n t_0}dt \cr &= \sum_{n,k=-N}^N p_k \delta_{k+n}
e^{-i n t_0} = \sum_{k=-N}^N p_k e^{i k t_0} = P(t_0)
\end{eqnarray}
3) $\delta_N(t)$, integrated with an arbitrary function, projects it
on the $span\{e^{- i N t}, e^{-i(N-1)t},...,e^{i N t}\}$ subspace
and acts like a real delta-function there. Indeed, consider the
function $f(t) = \sum_{k=-\infty}^{\infty} f_k e^{i k t}$. Then:
\begin{eqnarray}
\int f(t) \delta_N(t-t_0) dt &= \sum_{n=-N}^N {1\over 2\pi} \int
f(t)e^{i n t-i n t_0}dt \cr &=\sum_{n=-N}^N f_{-n} e^{-i n t_0} =
\sum_{n=-N}^N f_n e^{i n t_0}
\end{eqnarray}
Where the last expression is nothing but the function $f(t)$ which
higher modes have been cut-off. It is convenient to introduce the
following notation for this projection:\\
\textbf{Notation:}
\begin{equation}{
\label{projector} \left[f\right]_N(t) = \sum_{k=-N}^N f_k e^{i k t}
= \int f(t') \delta_N(t'-t) dt' }
\end{equation}
\subsubsection{Moyal product revisited}
Now we want to show that the correlator $^0\langle f_1(\phi(t_1))
f_2(\phi(t_2))\rangle$ gives rise to the famous Moyal product as
mentioned in the introduction. But first we should note that
$^0\langle\phi^i(t)\rangle=x^i$ and as
$^0\langle(\phi^i(t)-x^i)\rangle=0$:
\\
\textbf{Observation:} Wick theorem is held only for monomials over
the variables $\phi^i(t)-x^i$, e.g.:
 \begin{eqnarray}
 \label{wick_theorem}
^0\langle(\phi^{i_1}(t_1)-x^{i_1})(\phi^{i_2}(t_2)-x^{i_2})(\phi^{i_3}(t_3)-x^{i_3})(\phi^{i_4}(t_4)-x^{i_4})\rangle\cr
= \
^0\langle(\phi^{i_1}(t_1)-x^{i_1})(\phi^{i_2}(t_2)-x^{i_2})\rangle\
^0\langle(\phi^{i_3}(t_3)-x^{i_3})(\phi^{i_4}(t_4)-x^{i_4})\rangle+\cr
 +\ ^0\langle(\phi^{i_1}(t_1)-x^{i_1})(\phi^{i_3}(t_3)-x^{i_3})\rangle\ ^0\langle(\phi^{i_2}(t_2)-x^{i_2})(\phi^{i_4}(t_4)-x^{i_4})\rangle
 \end{eqnarray}
 As soon as $G(t,t)=0$ and $^0\langle(\phi^i(t_1)-x^i)(\phi^j(t_2)-x^j)\rangle=i\hbar\theta_{(0)}^{ij}G(t_1,t_2)$, we have $^0\langle(\phi^i(t)-x^i)(\phi^j(t)-x^j)\rangle=0$ and thus it is straightforward to show that $^0\langle f(\phi(t))\rangle = f(x)$. After this comment the derivation of the Moyal product becomes trivial - we provide a Taylor expansion around
 $\phi^i=x^i$:
 $$
 ^0\left\langle f_1(\phi(t_1)) f_2(\phi(t_2)) \right\rangle = f_1(x) f_2(x) + \partial_i f_1(x) \partial_j f_2(x)\ ^0\left\langle(\phi^i(t_1)-x^i)(\phi^j(t_2)-x^j)\right\rangle+
 $$
$$
+ {1\over(2!)^2}\partial_{i_1} \partial_{i_2} f_1(x)
\partial_{j_1}\partial_{j_2} f_2(x)\
^0\left\langle(\phi^{i_1}(t_1)-x^{i_1})(\phi^{i_2}(t_1)-x^{i_2})(\phi^{j_1}(t_2)-x^{j_1})(\phi^{j_2}(t_2)-x^{j_2})\right\rangle+\ldots
$$
$$
=f_1(x) f_2(x) + \partial_i f_1(x) \partial_j f_2(x)\
^0\left\langle(\phi^i(t_1)-x^i)(\phi^j(t_2)-x^j)\right\rangle+
$$
$$
+ {1\over 2!}\partial_{i_1} \partial_{i_2} f_1(x)
\partial_{j_1}\partial_{j_2}
f_2(x)\
^0\left\langle(\phi^{i_1}(t_1)-x^{i_1})(\phi^{j_1}(t_2)-x^{j_1})\right\rangle\
^0\left\langle(\phi^{i_2}(t_1)-x^{i_2})(\phi^{j_2}(t_2)-x^{j_2})\right\rangle+\ldots
$$
$$
=f_1(x) f_2(x) + \partial_i f_1(x) \partial_j f_2(x)
i\hbar\theta_{(0)}^{ij}G(t_1,t_2)+
$$
$$
+ {1\over 2!}\partial_{i_1} \partial_{i_2} f_1(x)
\partial_{j_1}\partial_{j_2} f_2(x)i\hbar\theta_{(0)}^{i_1 j_1}
G(t_1,t_2) i\hbar \theta_{(0)}^{i_2 j_2} G(t_1,t_2)+\ldots
$$
\begin{equation}{
\label{moyal_derivation} =f_1(x)
\exp{\left(i\hbar\overleftarrow{\partial_i}\theta_{(0)}^{ij}\overrightarrow{\partial_j}G(t_1,t_2)\right)}f_2(x)
  }
\end{equation}
If $t_1>t_2$ then
\begin{equation}{
\label{moyal_derived} ^0\left\langle f_1(\phi(t_1)) f_2(\phi(t_2))
\right\rangle = f_1(x) \exp{\left( {i\hbar\over
2}\overleftarrow{\partial_i}\theta_{(0)}^{ij}\overrightarrow{\partial_j}
\right)}f_2(x) }
\end{equation}
which is just the Moyal product.
\subsection{Perturbation: the first order in $\hbar$}
It is well-known that the tree level of the perturbation theory is
of the first order in $\hbar$ and reproduces the Poisson structure.
However there exist loop diagrams that could contribute to the
$O(\hbar)$ part of correlators and could spoil this property. These
loop contributions would definitely fall out of the commutators, as
the commutators \emph{have to} give the Poisson structure in the
first order in $\hbar$. In this subsection we check that such loop
contributions vanish in the $O(\hbar)$ and that in the first order
in $\hbar$ correlators give rise exactly to the Poisson structure.
To figure out this we act as follows. Consider:
\begin{equation}{
\label{first_order_h} {^0\left\langle f_1(\phi_N(t_1))
f_2(\phi_N(t_2)) \exp{\left({i\over\hbar}\int \left\{
e_i(\phi_N)\dot\phi_N^i + (de)_{ij}\psi_N^i\psi_N^j
\right\}dt\right)}\right\rangle_{N}\over
^0\left\langle\exp{\left({i\over\hbar}\int \left\{
e_i(\phi_N)\dot\phi_N^i + (de)_{ij}\psi_N^i\psi_N^j
\right\}dt\right)}\right\rangle_{N}} }
\end{equation}
\textbf{Observation:} We have the following:
\begin{equation}{
\label{singlepoint_O_h} {^0\left\langle \left(\phi_N^i(t) - x^i
\right) \exp{\left({i\over\hbar}\int \left\{ e_i(\phi_N)\dot\phi_N^i
+ (de)_{ij}\psi_N^i\psi_N^j \right\}dt\right)}\right\rangle_{N}\over
^0\left\langle\exp{\left({i\over\hbar}\int \left\{
e_i(\phi_N)\dot\phi_N^i + (de)_{ij}\psi_N^i\psi_N^j
\right\}dt\right)}\right\rangle_{N}}=O(\hbar^2) }
\end{equation}
The absence of the $O(1)$ term is obvious - the only such contribution is $^0\left\langle\left(\phi_N^i(t) - x^i\right)\right\rangle_{N} = 0$.\\
$O(\hbar)$ contribution arises from the following graphs:
\begin{equation}{
\label{selfcontr_single_leg} \epsfig{file=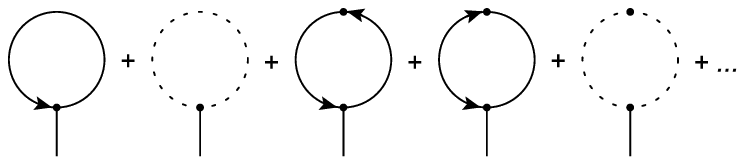} }
\end{equation} The appropriate analytical expression of the $n^{th}$
order contribution is proportional to:
\begin{eqnarray}
\label{singlepoint_vanishes} \theta_{(0)}^{ik}\theta_{(0)}^{j_1
i_2}\theta_{(0)}^{j_2 i_3}\ldots\theta_{(0)}^{j_n
i_1}\partial_k(de)_{i_1 j_1}(de)_{i_2 j_2}\ldots(de)_{i_n
j_n}\times\qquad\qquad\qquad\qquad\qquad\qquad\cr \times\int dt_1
\ldots dt_n \biggl[\Big(\delta_N(t_1-t_2)-\delta_N(t_2-\pi)
\Big)\Big(\delta_N(t_2-t_3)-\delta_N(t_3-\pi)\Big)\ldots\vphantom{\int}
\times\qquad\cr
\times\Big(\delta_N(t_n-t_1)-\delta_N(t_1-\pi)\Big)-\delta_N(t_1-t_2)\delta_N(t_2-t_3)\dots\delta_N(t_n-t_1)
\vphantom{\int}\biggr]G_N(t,t_1)\qquad
\end{eqnarray}
and vanishes due to $G_N(t,\pi)=0$ (as soon as $G_N(t,t')$ as a
function of $t$ has the support $\{-N,-N+1,...,N-1,N\}$ in the
Fourier space, we can take into account the second property of
$\delta_N(t)$ and work with $\delta_N(t)$ as with the standart
delta-function in ($\ref{singlepoint_vanishes}$)). Now consider:
\begin{equation}{
\label{twopoint} {^0\left\langle \left(\phi_N^i(t_1) - x^i \right)
\left(\phi_N^j(t_2) - x^j \right) \exp{\left({i\over\hbar}\int
\left\{ e_i(\phi_N)\dot\phi_N^i + (de)_{ij}\psi_N^i\psi_N^j
\right\}dt\right)}\right\rangle_{N}\over
^0\left\langle\exp{\left({i\over\hbar}\int \left\{
e_i(\phi_N)\dot\phi_N^i + (de)_{ij}\psi_N^i\psi_N^j
\right\}dt\right)}\right\rangle_{N}} }
\end{equation}
The $O(1)$ term is absent again, while the $O(\hbar)$ term is given
by the sum over the tree diagrams:
$$
\epsfig{file=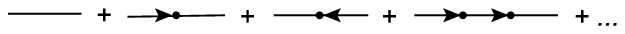}
$$
We should notice that every \epsfig{file=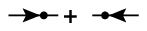} pair gives
rise to the $de(x)$ factor after integrating by parts. So the
appropriate analytical expression is:
\begin{eqnarray}
\label{twopoint_computation} i\hbar\theta_{(0)}^{ij}G(t_1,t_2) +
i^3\hbar\theta_{(0)}^{i
i_1}\big(\partial_{i_1}e_{j_1}(x)-\partial_{j_1}e_{i_1}(x)\big)\theta_{(0)}^{j_1
j}\int G(t_1,t){d G(t,t_2)\over dt} dt+\cr + i^5\hbar\theta_{(0)}^{i
i_1}(de)_{i_1 j_1}\theta_{(0)}^{j_1 i_2}(de)_{i_2
j_2}\theta_{(0)}^{j_2 j}\int G(t_1,t){d G(t,t')\over dt}{d
G(t',t_2)\over dt'}dt dt' +\ldots\cr =i\hbar G(t_1,t_2)\left(
\theta_{(0)} - \theta_{(0)} de \theta_{(0)} + \theta_{(0)} de
\theta_{(0)} de \theta_{(0)} -... \right)^{ij} = i\hbar G(t_1,t_2)
\theta^{ij}
\end{eqnarray}
where $\theta=\left(\omega^{(0)} + de \right)^{-1}$ and we omitted
the obvious index notations inside the last brackets. We used
($\ref{derivative_of_prop}$) here. Notice that we took off the
regularization in ($\ref{twopoint_computation}$), i.e. put
$N=\infty$ - it is acceptable at the tree level as it is well known
in QFT. Taking into account ($\ref{twopoint_computation}$) and that
$G_N(t,t)=0$ we can conclude that ($\ref{twopoint}$) vanishes in the
case $t_1 = t_2 = t$.\\
One more thing to mention is that higher degree monomials in
$\phi^i(t)-x^i$ (higher than 2) giver rise to the $O(\hbar^2)$ terms
and thus can be neglected.\\ Now it's easy to see that:
\begin{eqnarray}
\label{first_order_computation} {^0\left\langle f_1(\phi(t_1))
f_2(\phi(t_2)) \exp{\left({i\over\hbar}\int \left\{
e_i(\phi)\dot\phi^i + (de)_{ij}\psi^i\psi^j
\right\}dt\right)}\right\rangle\over
^0\left\langle\exp{\left({i\over\hbar}\int \left\{
e_i(\phi)\dot\phi^i + (de)_{ij}\psi^i\psi^j
\right\}dt\right)}\right\rangle}=f_1(x)f_2(x) +\qquad\qquad\cr +
\partial_i f_1(x)\partial_j f_2(x){^0\left\langle \left(\phi^i(t_1)
- x^i \right) \left(\phi^j(t_2) - x^j \right)
\exp{\left({i\over\hbar}\int \left\{ e_i(\phi)\dot\phi^i +
(de)_{ij}\psi^i\psi^j \right\}dt\right)}\right\rangle\over
^0\left\langle\exp{\left({i\over\hbar}\int \left\{
e_i(\phi)\dot\phi^i + (de)_{ij}\psi^i\psi^j
\right\}dt\right)}\right\rangle} +\cr+ O(\hbar^2) = f_1(x)f_2(x) +
{i\hbar\over 2}\theta^{ij}\partial_i f_1(x) \partial_j f_2(x) +
O(\hbar^2)\qquad\qquad
\end{eqnarray}
where we have assumed that $t_1>t_2$. As predicted, the $O(\hbar)$
term of the correlator (the tree level term) is nothing but the
Poisson structure.
\subsection{Loop calculations}
In this subsection we will compute the simplest loop diagram to
demonstrate some special features of the subject.
\subsubsection{Anomalous vertex: first order in $e$}
In the first order in perturbation theory loops can be obtained only
by self-contractions. As long as $^0\langle
\left(\phi_N^i(t)-x^i\right)\left(\phi_N^j(t)-x^j\right)\rangle_N=0$,
only $^0\langle
\left(\phi_N^i(t)-x^i\right)\dot\phi_N^j(t)\rangle_N$ contraction
contributes. Therefore we are left with the following diagrams:
$$
\epsfig{file=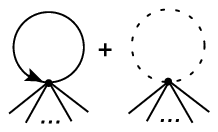}
$$
We will think of this as of a part of the bigger diagram. The
analytical expression is:
\begin{eqnarray}
\label{pert_first_order} {1\over n!}{i\over\hbar}\int
dt\left(\phi_N^{i_1}(t)-x^{i_1}\right)...\left(\phi_N^{i_n}(t)-x^{i_n}\right)\bigg(
\partial_{i_1}...\partial_{i_n}\partial_i
e_j(x)\ ^0\langle\phi_N^i(t)\dot\phi_N^j(t)\rangle_N +\cr
+\partial_{i_1}...\partial_{i_n}\left(de(x)\right)_{ij}\
^0\langle\psi_N^i(t)\psi_N^j(t)\rangle_N \bigg)
\end{eqnarray}
$^0\langle\phi^i\dot\phi^j\rangle$ is proportional to $\theta^{ij}$
and therefor is antisymmetric in $i, j$. So we can rewrite
($\ref{pert_first_order}$) as follows:
\begin{eqnarray}
\label{pert_first_rewrite} {1\over n!}{i\over\hbar}\int
dt\left(\phi_N^{i_1}(t)-x^{i_1}\right)...\left(\phi_N^{i_n}(t)-x^{i_n}\right)\partial_{i_1}...\partial_{i_n}\left(de(x)\right)_{ij}\bigg[{1\over
2}\ ^0\langle\phi_N^i(t)\dot\phi_N^j(t)\rangle_N +\cr + \
^0\langle\psi_N^i(t)\psi_N^j(t)\rangle_N \bigg]
\end{eqnarray}
The expression inside the square brackets is:
\begin{eqnarray}
\label{loop_sum} &{1\over 2}\
^0\langle\phi_N^i(t)\dot\phi_N^j(t)\rangle_{N} + \
^0\langle\psi_N^i(t)\psi_N^j(t)\rangle_{N}&={i\hbar\over
2}\theta_{(0)}^{ij}\left[ \delta_N(t-\pi) - \delta_N(0) +
\delta_N(0) \right]\cr&&={i\hbar\over 2}\theta_{(0)}^{ij}
\delta_N(t-\pi)
\end{eqnarray}
where $\delta_N(t)$ is a mode-regularized $\delta$-function
($\ref{regular_delta}$) and we have used
($\ref{derivative_of_prop}$). Substituting ($\ref{loop_sum}$) into
($\ref{pert_first_rewrite}$) we get an effective n-boson vertex:
\begin{equation}{
\label{anom_vert} V_n=-{1\over 2}{1\over n!}\int
dt\left(\phi_N^{i_1}(t)-x^{i_1}\right)...\left(\phi_N^{i_n}(t)-x^{i_n}\right)\partial_{i_1}...\partial_{i_n}\left(de(x)\right)_{ij}\theta^{ij}_{(0)}\delta_N(t-\pi)\
\ =\ \ \epsfig{file=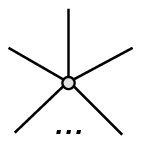} }
\end{equation}
we can rewrite it using our projector:
\begin{equation}{
\label{anom_vert_comp} V_n = -{1\over 2}{1\over
n!}\partial_{i_1}...\partial_{i_n}\left(de(x)\right)_{ij}\theta^{ij}_{(0)}\left[\left(\phi_N^{i_1}-x^{i_1}\right)...\left(\phi_N^{i_n}-x^{i_n}\right)
\right]_N(\pi) }
\end{equation}
and then notice that:
\begin{equation}{
\label{anom_vert_big} V = \sum_{n=0}^{\infty} V_n = -{1\over
2}\theta^{ij}_{(0)}\left[de_{ij} \right]_N(\pi) }
\end{equation}
which is the claimed anomalous vertex as we'll see later. The contribution depends on the number of legs $n$.\\
\textbf{Single leg case}\\
In a case $n=1$ ($\ref{anom_vert_comp}$) gives rise to the
following:
\begin{equation}{
\label{singleleg} V_1=-{1\over
2}\partial_{i_1}\left(de(x)\right)_{ij}\theta^{ij}_{(0)}\left[\phi_N^{i_1}-x^{i_1}\right]_N(\pi)
= -{1\over
2}\partial_{i_1}\left(de(x)\right)_{ij}\theta_{(0)}^{ij}\left(
\phi^{i_1}_N(\pi)-x^{i_1} \right)=0 }
\end{equation}
Here we used:
\begin{equation}{
\label{int_with_delta_reg} \left[\phi^i_N\right]_N(t)=\phi^i_N(t) }
\end{equation}
The last expression in ($\ref{singleleg}$) vanishes due to the
delta-function fixing $\phi^i_N(\pi)=x^i$.

We could also contract this single leg to the external field
$\phi^a(t_1)$, then the left-hand side of ($\ref{singleleg}$) would
be proportional to
\begin{equation}{
\label{int_with_delta_again} \int G_N(t_1,t)\delta_N(t-\pi)
dt=G_N(t_1,\pi)=0 }
\end{equation}
Notice that although expression $\int G(t_1, t)\delta(t-\pi) dt$ is
ill-defined ($G(t_1,t)$ is discontinuous at $t=\pi$), we still can
make sense of it and say $\int G_N(t_1,t)\delta_N(t-\pi)
dt=G_N(t_1,\pi)=0$ in a mode-regularization sense - we've already
used this in the section 3.2 when explaining that the diagrams
($\ref{selfcontr_single_leg}$) didn't contribute. The situation will
be pretty different with two or more legs - the delta-function
$\delta_N(t)$ will not contain enough modes to make naive identities
like $\int G_N(t_1,t)G_N(t_2,t)\delta_N(t-t')dt =
G_N(t_1,t')G_N(t_2,t')$ correct.\\
\textbf{Two and more legs case}\\
Consider
\begin{equation}{
\label{twolegs} V_2 = -{1\over 4}\int
dt\left(\phi_N^{i_1}(t)-x^{i_1}\right)\left(\phi_N^{i_2}(t)-x^{i_2}\right)\partial_{i_1}\partial_{i_2}\left(de(x)\right)_{ij}\theta_{(0)}^{ij}
\delta_N(t-\pi) }
\end{equation}
which is no longer zero because in the Fourier space the support of
$\left(\phi_N^{i_1}(t)-x^{i_1}\right)\left(\phi_N^{i_2}(t)-x^{i_2}\right)$
is $\{-2N,...,2N\}$.

To see this in a different manner let us contract the two exterior
legs from ($\ref{twolegs}$) with exterior fields $\phi_N^a(t_1)$ and
$\phi_N^b(t_2)$:
$$
\epsfig{file=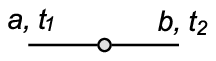}
$$
This gives rise to:
\begin{equation}{
\label{twolegs_answer_calculate} -{(i\hbar)^2\over 2}\theta_{(0)}^{a
i_1}\theta_{(0)}^{b i_2}\theta_{(0)}^{ij}
\partial_{i_1}\partial_{i_2}\left(de(x)\right)_{ij}\int G_N(t_a,t)
G_N(t_b,t) \delta_N(t-\pi) dt }
\end{equation}
Introduce a notation\footnote{In the n-legs case we'll obtain the
similar integral:
$$
\label{definition_of_an}
A^{(n)}(t_1,t_2,...,t_n)=\lim_{N\to\infty}\int G_N(t_1,t)
G_N(t_2,t)...G_N(t_n,t) \delta_N(t-\pi) dt $$}:
\begin{equation}{
\label{definition_of_a2}
A^{(2)}(t_a,t_b)=\lim_{N\to\infty}A_N^{(2)}(t_a,t_b)=\lim_{N\to\infty}\int
G_N(t_a,t) G_N(t_b,t) \delta_N(t-\pi) dt }
\end{equation}
The statement that ($\ref{twolegs}$) really contributes means that
$A^{(2)}\ne 0$ and we finally get:
\begin{equation}{
\label{twolegs_answer} ^0\langle \phi^a_N(t_1) V_2 \phi^b_N(t_2)
\rangle={\hbar^2\over 2}\theta_{(0)}^{a i_1}\theta_{(0)}^{b
i_2}\theta_{(0)}^{ij}
\partial_{i_1}\partial_{i_2}\left(de(x)\right)_{ij}A^{(2)}(t_a,t_b)
}
\end{equation}
This answer will be crucial for us as it will describe the lowest anomaly in our theory.\\
\subsubsection{Nontrivial integral}

We introduced the function $A^{(2)}(t_1,t_2)$ above. Understanding
of it's behavior is important to understand the anomaly that will be
discussed in the last section of the paper.

Let us demonstrate some special issues of this function. One could
naively think according to ($\ref{definition_of_a2}$) that the limit
can be safely taken at the very beginning and that
$A^{(2)}(t_a,t_b)=\int G(t_a,t)G(t_b,t)\delta(t-\pi) dt$, but this
expression is ill-defined as $G(t',t)$ contains a jump at $t=\pi$.
To be more consistent let us consider the following:
\begin{equation}{
\label{play_with_a} A^{(2)}_{N_1, N_2}(t_a,t_b) = \int
G_{N_1}(t_a,t) G_{N_1}(t_b,t) \delta_{N_2}(t-\pi) dt }
\end{equation}
and observe how this behaves when $N_1, N_2 \to \infty$. From one
point:
\begin{eqnarray}
\label{limit_a} \lim_{^{N_1\to\infty}_{{N_2\over
N_1}\to\infty}}A^{(2)}_{N_1, N_2}(t_a,t_b) &= \lim_{N_1\to\infty}
\int G_{N_1}(t_a,t) G_{N_1}(t_b,t) \delta(t-\pi) dt \cr&=
\lim_{N_1\to\infty} G_{N_1}(t_a,\pi) G_{N_1}(t_b,\pi) =0
\end{eqnarray}
From another point of view we could do the following:
\begin{equation}{
\label{limit_b} \lim_{^{N_2\to\infty}_{{N_1\over
N_2}\to\infty}}A^{(2)}_{N_1, N_2}(t_a,t_b) = \lim_{N_2\to\infty}
\int G(t_a,t) G(t_b,t) \delta_{N_2}(t-\pi) dt}
\end{equation}
But we know the behavior of $G(t',t)$ in the vicinity of the point
$t=\pi$ - it is constant and equal to $-{1\over 2}$ to the left,
constant and equal to ${1\over 2}$ to the right and vanishes at
$t=\pi$. Thus $G(t_1,t)G(t_2,t)$ is constant and equal to ${1\over
4}$ at the vicinity of $t=\pi$ except of the point $t=\pi$ itself
where it vanishes. But integration with a smooth function
$\delta_{N_2}(t-\pi)$ is insensitive to this single discontinuity
and thus we conclude that:
\begin{equation}{
\label{limit_b_res} \lim_{^{N_2\to\infty}_{{N_1\over
N_2}\to\infty}}A^{(2)}_{N_1, N_2}(t_a,t_b) = {1\over 4}
 }
\end{equation}
So what have we got? The answer depends on how we take the limit and
thus we cannot put $N=\infty$ at the very beginning. We should
compute ($\ref{definition_of_a2}$) and take $N\to\infty$ at the end.

Let us do this in details. Substituting Fourier transforms of
propagators and delta-function into ($\ref{definition_of_a2}$) one
gets:
\begin{eqnarray}
\label{computation_of_a2}
A^{(2)}_N(t_1,t_2)=\sum_{n_1,m_1,n_2,m_2,p=-N}^N{1\over 2\pi}\int
G_{n_1 m_1}e^{i n_1 t_1 + i m_1 t} G_{n_2 m_2} e^{i n_2 t_2 + i m_2
t} e^{i p (t-\pi)} dt\cr =\sum_{n_1,m_1,n_2,m_2,p=-N}^N e^{i n_1 t_1
+ i n_2 t_2} G_{n_1 m_1} G_{n_2 m_2} (-1)^p\delta_{m_1 + m_2 +
p}\qquad\qquad\qquad\cr =\sum_{n_1,m_1,n_2,m_2=-N}^N e^{i n_1 t_1 +
i n_2 t_2} G_{n_1 m_1} G_{n_2 m_2} (-1)^{m_1+m_2}\delta(|m_1 +
m_2|\le N)\qquad
\end{eqnarray}
where we used the notation $\delta(true)=1$, $\delta(false)=0$ and
the fact that if $m_1+m_2+p=0$ for $|p|\le N$ then $|m_1+m_2|\le N$.
Substituting expression for $G_{nm}$ and providing computations we
get:
\begin{eqnarray}
\label{series_for_a2} A^{(2)}_N(t_1,t_2)=-{1\over
(2\pi)^2}\sum_{^{n,m=-N}_{n,m \ne 0}}^N{\delta(|n+m|\le N)\over n
m}\left(e^{i n (t_1 -\pi)}-1\right)\left(e^{i m (t_2
-\pi)}-1\right)+\cr + {1\over (2\pi)^2}\sum_{^{n,m=-N}_{n,m \ne
0}}^N {e^{i n(t_1 -\pi) + i m (t_2 - \pi)}\over n m}\qquad
\end{eqnarray}
It's difficult to study the limit $N\to\infty$ in the complete
expression ($\ref{series_for_a2}$) for $A^{(2)}$, but we can take a
roundabout and compute $A^{(2)}(0,0)$ and $\int A^{(2)}(t_1,t_2)dt_1
dt_2$ rather easily. Then we'll check the full answer numerically.
\begin{eqnarray}
\label{integral_of_a2} \int A_N^{(2)}(t_1,t_2) dt_1 dt_2 =
-\sum_{^{n,m=-N}_{n,m \ne 0}}^N{\delta(|n+m|\le N)\over n
m}=-\sum_{^{n,m=-N}_{n,m \ne 0}}^N\left[{1\over n m} -
{\delta(|n+m|> N)\over n m}\right]\cr =\sum_{^{n,m=-N}_{n,m \ne
0}}^N{\delta(|n+m|> N)\over n m}=2\sum_{(n,m)\in D_N}{1\over n
m}\qquad\qquad
\end{eqnarray}
Where $D_N$ is a region illustrated below:
$$
\epsfig{file=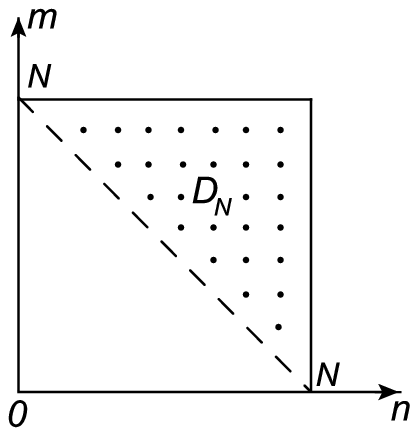}
$$
\begin{equation}{
\label{int_sum1} \lim_{N\to\infty}2\sum_{(n,m)\in D_N}{1\over n
m}=\lim_{N\to\infty} 2\sum_{(p,q)\in D_N}{1\over {p\over N}{q\over
N}}{1\over N}{1\over N} = 2\int_{(x,y)\in D}{dx dy\over x y} }
\end{equation}
Where we we've replaced an integral sum by the appropriative
integral and the region $D$ is:
$$
\epsfig{file=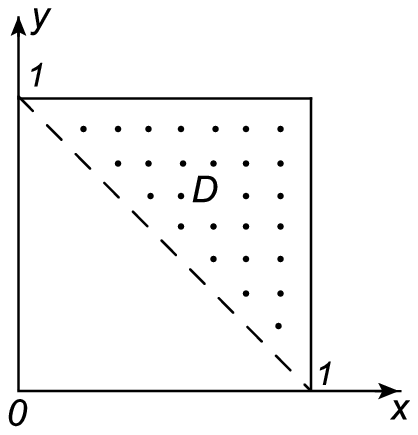}
$$
Then we can find:
\begin{equation}{
\label{int_of_a2_ans} \lim_{N\to\infty}\int A_N^{(2)}(t_1,t_2)dt_1
dt_2=2\int_0^1{dx\over x}\int_{1-x}^1{dy\over y}=-2\int_0^1 {dx\over
x}\ln(1-x)={\pi^2\over 3}
  }
\end{equation}
We can also find:
\begin{equation}{
\label{a2_at_zero} A_N^{(2)}(0,0)={1\over 2\pi^2}\sum_{^{(n,m)\in
D_N}_{n,m\ -\ odd}}{4\over n m}={1\over 2\pi^2}\sum_{^{(n,m)\in
D_N}_{n,m\ -\ odd}}{1\over {n\over N}{m\over N}}{2\over N}{2\over N}
}
\end{equation}
From where we see:
\begin{equation}{
\label{one_twelfth} \lim_{N\to\infty} A^{(2)}_N(0,0)={1\over
2\pi^2}\int_{(x,y)\in D} {dx dy\over xy}={1\over 12} }
\end{equation}
Comparing ($\ref{one_twelfth}$) and ($\ref{int_of_a2_ans}$) and
noticing that ${(2\pi)^2\over 12}={\pi^2\over 3}$ one could suppose
that $A(t_1,t_2)$ is equal to ${1\over 12}$ everywhere except of
$t_1=\pi$ or $t_2=\pi$ where it vanishes. Such prediction has been
checked numerically and turns out to be correct. So we conclude
that:
\begin{equation}{
\label{value_of_a2} A^{(2)}(t_1,t_2) = \begin{cases}{1\over 12},&if\
t_1\ne\pi\ {\rm and}\ t_2\ne\pi;\cr 0, &otherwise.\cr\end{cases} }
\end{equation}
What is interesting is that the obtained value lies between the two
naive predictions: $0<{1\over 12}<{1\over 4}$.
\subsection{Conclusion to the section}
In this section we provided illustrative examples of path integral approach for the phase space. The experience obtained is that all the quantities are well-defined at the tree level, i.e. up to the first order in $\hbar$. At this point the answer happens to be background-independent and the Poisson structure is really restored as expected. However the situation becomes more complicated if we take the loops into account. Naive arguments don't make sense any more - we have to work carefully with respect to the regularization.\\
We can also notice that ($\ref{twolegs_answer}$) deforms the Moyal
star-product:
\begin{eqnarray}
\label{deformed_prod} \langle f_1(\phi(t_1)) f_2(\phi(t_2))\rangle
=\ ^0\langle f_1(\phi(t_1)) f_2(\phi(t_2))\rangle
+\qquad\qquad\qquad\qquad\qquad\qquad\qquad\qquad\cr+ {\hbar^2\over
24}\theta_{(0)}^{a i_1}\theta_{(0)}^{b
i_2}\theta_{(0)}^{ij}\partial_{i_1}\partial_{i_2}(de(x))_{ij}\partial_a
f_1(x)\partial_b f_2(x) + o(e) + o(\hbar^2)\qquad
\end{eqnarray}
which is obviously non-invariant under diffeomorphisms and indicates
that there is some kind of anomaly present here -- that is the topic
of the next section.
\section{Anomaly chasing}
In this section we are going to analyze our theory behavior with
respect to the target-space \emph{nonlinear} diffeomorphisms. We
will show that it is not invariant due to the UV cut-off being
non-invariant under such diffeomorphisms.

For the purpose of simplicity we will put $x^i=0$ when providing
concrete calculations starting from the subsection 4.2, i.e. all the
fields will vanish at $t=\pi$. We shall also assume throughout this
section that $e_i(\varphi)$ is a real analytical function whose
power series starts from the quadratic term. The same assumption
will be about $v^i(\varphi)$ (the vector field describing our
infinitesimal diffeomorphism - to be introduced soon).

There are two ways to act with diffeomorphism in a theory with
regularization: we either provide naive continuous diffeomorphism
and then regularization or provide regularization first and then
make diffeomorphism in the finite-dimensional space of regularized
fields; this finite-dimensional diffeomorphism imitates the real one
in the $N\to\infty$ limit. We'll refer to the first prescription as
to the naive one and to the second - as to the proper one.
\subsection{Naive prescription}
We provide a classical diffeomorphism first:
\begin{equation}{
\label{diff_boson_s1} \phi^i = \varphi^i + \epsilon v^i(\varphi) }
\end{equation}
\begin{equation}{
\label{diff_fermion_s1} \psi^i = \widetilde{\psi}^i + \epsilon
\partial_j v^i(\varphi) \widetilde{\psi}^j }
\end{equation}
and study corresponding Ward identities.\\
We make the following naive assumptions:

1) Supersymmetric measure is invariant - a rather natural
assumption.

2) Delta-function regulator is invariant. This should be explained.
The delta-function transforms as follows:
$$
\delta\left(\phi(\pi)-x\right) = \delta\left(\varphi(\pi) + \epsilon
v(\varphi(\pi)) - y - \epsilon v(y)\right)=
$$
\begin{equation}{
\label{delta_transform} ={\delta\left(\varphi(\pi) - y\right)\over
|det\left[\delta^i_j + \epsilon\partial_j
v^i\left(\varphi(\pi)\right)\right]|}={\delta\left(\varphi(\pi) -
y\right)\over |det\left[\delta^i_j + \epsilon\partial_j
v^i\left(y\right)\right]|} }
\end{equation}
where $x^i = y^i + \epsilon v^i(y)$. We see that the ${1\over
|\det(...)|}$ multiplier is naively constant as delta-function
guarantees $\varphi^i(\pi)=y^i$, hence it is canceled out from the
correlators and we can think of the delta-function as of an
invariant object at this point\footnote{However when we provide a
proper approach, we'll see that the last equality in
($\ref{delta_transform}$) will not be true. The case is that
delta-function non-invariance ${1\over \left|det\left[\delta^i_j +
\epsilon\partial_j v^i\left(\phi(\pi)\right)\right]\right|}$ under
correlators will give rise to expressions like
$\langle\varphi(t)\varphi(\pi)\rangle\langle\varphi(t')\varphi(\pi)\rangle\delta(\varphi(\pi))$
that are \emph{naively} zero but in fact do not vanish as we have
seen in the ``Anomalous vertex'' subsection -- because propagator
contains jump at $t=\pi$.}.

3) The action transforms in a classical way:
\begin{equation}{
\label{act_transform} S=\int \left\{{1\over
2}\omega^{(0)}_{ij}\varphi^i\dot\varphi^j +
e_i(\varphi)\dot\varphi^i +
\epsilon\omega^{(0)}_{ij}v^i(\varphi)\dot\varphi^j +
\epsilon(\mathcal{L}_v e)_i\dot\varphi^i \right\}dt+ S_f }
\end{equation}
Where the fermionic part of the action:
\begin{equation}{
\label{act_trans_fermions} S_f = \int \left\{
\omega^{(0)}_{ij}\widetilde{\psi}^i\widetilde{\psi}^j +
(de)_{ij}\widetilde{\psi}^i\widetilde{\psi}^j +
2\epsilon\omega^{(0)}_{kj}\partial_i v^k
\widetilde{\psi}^i\widetilde{\psi}^j + \epsilon(\mathcal{L}_v
de)_{ij}\widetilde{\psi}^i\widetilde{\psi}^j  \right\} }
\end{equation}
where $\mathcal{L}_v$ is a Lie derivative with respect to the vector field $v$ along the diffeomorphism.\\
4) The inserted observables also transform classically: $F[\phi]\to F[\varphi] + \epsilon\int {\delta F[\varphi]\over \delta \varphi^i}v^i$.\\

At the end of the day we provide mode-regularization, namely use
($\ref{fourier_reg}$). After all these procedures have been done we
extract the first order part in $\epsilon$ and assume it to vanish
(if the invariance is not broken) -- that is the way we get the Ward
identities. If the identity does not hold -- we get an anomaly.

\subsubsection{Ward identities and their breakdown}
We start with a general interacting theory and are interested in a
two-point correlation function:
\begin{equation}{
\label{undeformed_corr}
\langle\phi^a(t_1)\phi^b(t_2)\rangle={\int\mathcal{D}\phi\mathcal{D}\psi
\phi^a(t_1)\phi^b(t_2) e^{{i\over\hbar}\int\left\{{1\over
2}\omega^{(0)}_{ij}\phi^i\dot\phi^j + e_i(\phi)\dot\phi^i +
ferm.\right\}}\delta\left(\phi^i(\pi)\right)\over
\int\mathcal{D}\phi\mathcal{D}\psi
e^{{i\over\hbar}\int\left\{{1\over
2}\omega^{(0)}_{ij}\phi^i\dot\phi^j + e_i(\phi)\dot\phi^i +
ferm.\right\}}\delta\left(\phi^i(\pi)\right)} }
\end{equation}
and provide the classical diffeomorphism
($\ref{diff_boson_s1}$)-($\ref{diff_fermion_s1}$). After that the
action transforms into ($\ref{act_transform}$), the measure remains
the same, as well as the delta-function (as explained above).
Observables transform as follows:
\begin{equation}{
\label{obs_transform} \phi^a(t_1)\phi^b(t_2) = \varphi^a(t_1)
\varphi^b(t_2) +  \epsilon \varphi^a(t_1)
v^b\left(\varphi(t_2)\right) + \epsilon
v^a\left(\varphi(t_1)\right)\varphi^b(t_2) + o(\epsilon) }
\end{equation}
Then we extract the first order in $\epsilon$ and assume it to
vanish as long as we naively await that coordinate change does not
affect the result. Namely the following:
\begin{equation}{
\label{transformed_corr}
{\left\langle\varphi^a(t_1)\varphi^b(t_2)\left(1 + \epsilon
{i\over\hbar}\int \left\{ \omega^{(0)}_{ij}v^i\dot\varphi^j +
(\mathcal{L}_v e)_i \dot\varphi^i + fermions \right\}dt \right) +
\epsilon\left( \varphi^a v^b + v^a\varphi^b
\right)\right\rangle\over 1 + \epsilon\left\langle {i\over\hbar}\int
\left\{ \omega^{(0)}_{ij}v^i\dot\varphi^j + (\mathcal{L}_v e)_i
\dot\varphi^i + fermions \right\}dt \right\rangle} }
\end{equation}
should coincide with $\langle\phi^a(t_1)\phi^b(t_2)\rangle$. That is
the way to get the Ward identity:
\begin{eqnarray}
\label{naive_ward_id} \left\langle \varphi^a(t_1)
v^b\left(\varphi(t_2)\right)\right\rangle + \left\langle
v^a\left(\varphi(t_1)\right)\varphi^b(t_2)  \right\rangle
+\qquad\qquad\qquad\qquad\qquad\qquad\qquad\cr
 + \left\langle \varphi^a(t_1)\varphi^b(t_2) {i\over\hbar}\int \left\{ \omega^{(0)}_{ij}v^i\dot\varphi^j + (\mathcal{L}_v e)_i \dot\varphi^i + fermions \right\}dt
 \right\rangle_{\widehat{ab}}\begin{array}{c}
_{naively}\\
=\\
\\
\end{array}0
\end{eqnarray}
where the lower index "$\widehat{ab}$" in the last term means that
we should not contract $\phi^a(t_1)$ with $\phi^b(t_2)$ - the
appropriate term is canceled out by the denominator in
($\ref{transformed_corr}$).

All the propagators of our theory are defined in a perturbative
manner. That means that ($\ref{naive_ward_id}$) is not just a single
identity - it encodes the whole series of identities written in
terms of the correlators of quadratic theory $\langle...\rangle_0$:
the zeroth order (in $e(\varphi)$) identity, the first order
identity and so on. We will need only the zeroth order one for our
purposes. Let us extract it explicitly:
\begin{eqnarray}
\label{zeroth_naive_id} \ ^0\langle \varphi^a v^b\rangle +\
^0\langle v^a \varphi^b\rangle + {i\over\hbar}\int\ ^0\bigg\langle
\varphi^a(t_1) \varphi^b(t_2)
\bigg\{\omega^{(0)}_{ij}v^i\big(\varphi(t)\big)\dot\varphi^j(t)
+\qquad\qquad\qquad\qquad\cr +2\omega^{(0)}_{k j}\partial_i
v^k\big(\varphi(t)\big)\widetilde{\psi}^i(t)\widetilde{\psi}^j(t)
\bigg\} \bigg\rangle_{\widehat{ab}} dt \begin{array}{c}
_{naively}\\
=\\
\\
\end{array}0\qquad
\end{eqnarray}
where index "$\widehat{ab}$" again disallows contraction between
$\varphi^a$ and $\varphi^b$. We will examine this identity below.
\subsubsection{The zeroth order Ward identity}
We already stated in the introduction to the chapter that we were
interested in nonlinear transformations and that $v^i(\varphi)$
(being thought of as a real analytical function) was a power series
starting with quadratic terms: $v^i(\varphi) = {1\over
2}\partial_a\partial_b v^i(0) \varphi^a\varphi^b + ...$. Analogously
we've stated that $e_i(\varphi) = {1\over 2}\partial_a\partial_b
e_i(0)\varphi^a\varphi^b + ...$. Taking these into account we
conclude that the first two terms in ($\ref{zeroth_naive_id}$)
vanish as long as there is either odd number of fields inside the
correlator or self-contractions of $v^i$ (which do vanish) present.
So we are left with the third term.

If we compute it explicitly we will get an answer that will be a
formal power series in $\hbar$. Then vanishing of it means that
every coefficient of this formal series vanishes. It's easy to see
that this series starts with the $O(\hbar^2)$ - we will calculate
only such term. One can see (by observing that there should be 6
fields inside the correlator) that $O(\hbar^2)$ term comes from the
cubic part of $v^i(\varphi)$. The only possible contraction in such
term (that do not contain self-contractions of $v^i$ which vanish)
looks as follows:
\begin{eqnarray}
\label{naive_zeroth} {i\over\hbar}\int\ ^0\left\langle
\varphi^a(t_1)\varphi^l(t)\right\rangle\
^0\left\langle\varphi^b(t_2) \varphi^s(t)\right\rangle
\bigg\{\omega^{(0)}_{ij}\partial_l\partial_s \partial_k v^i(0)\
^0\langle\varphi^k(t)\dot\varphi^j(t)\rangle +\cr + 2\omega^{(0)}_{k
j}\partial_l\partial_s\partial_i v^k(0)\
^0\langle\widetilde{\psi}^i(t)\widetilde{\psi}^j(t)\rangle \bigg\}
dt
\end{eqnarray}

But it is easy to notice that this expression describes nothing but
the two-legs case with self-contraction considered in the
``Anomalous vertex'' subsection, namely it corresponds to the
diagram:
$$
\epsfig{file=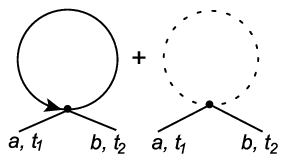}
$$

So to write an answer for ($\ref{naive_zeroth}$) one may just use
the result ($\ref{twolegs_answer}$) and make the substitution
$(de)_{ij} \to d (\omega^{(0)}_{kp}v^k d\varphi^p)_{ij} =
\omega^{(0)}_{kj}\partial_i v^k - \omega^{(0)}_{ki}\partial_j v^k$
in it. So we get:
\begin{eqnarray}
\label{lowest_anom_contr} {\hbar^2\over 2}\theta_{(0)}^{a
i_1}\theta_{(0)}^{b i_2}\theta_{(0)}^{ij}
\partial_{i_1}\partial_{i_2}\left(\omega^{(0)}_{kj}\partial_i v^k(0)
- \omega^{(0)}_{ki}\partial_j v^k(0)\right)A^{(2)}(t_1,t_2)\cr
=-\hbar^2\theta_{(0)}^{a i_1}\theta_{(0)}^{b i_2}
\partial_{i_1}\partial_{i_2}\partial_i v^i(0) A^{(2)}(t_1,t_2)
\end{eqnarray}
where we used $\omega^{(0)}_{ij}\theta_{(0)}^{jk}=\delta_i^k$. But
we know that $A^{(2)}(t_1,t_2)\ne 0$. So
\begin{equation}{
\label{ward_id_breakdown} -\hbar^2\theta_{(0)}^{a
i_1}\theta_{(0)}^{b i_2}
\partial_{i_1}\partial_{i_2}\partial_i v^i(0) A^{(2)}(t_1,t_2)\ne 0
}
\end{equation}
and finally we see that the naive Ward identity is not valid.
\subsection{Proper prescription}
However we could act differently. We could go to the regularized
fields ($\ref{fourier_reg}$) first and then provide diffeomorphism
in terms of their Fourier modes. If we put naively:
\begin{equation}{
\label{naive_reg_diff} \phi^i_N(t) = \varphi^i_N(t) + \epsilon
v^i(\varphi_N(t)) }
\end{equation}
then we'll see that $\phi_N$ and $\varphi_N$ connected in such a way
can't have an equal support in the Fourier space. Indeed, if the
support of $\varphi^i_N$ is $\{-N,...,N\}$ then $\phi^i_N$ will
necessarily contain modes higher than $N$ for the case of general
diffeomorphism. Only linear transformations do not generate higher
modes but if we are interested in a non-linear transformation (and
in fact we are) we will always get these higher modes.

The most straightforward way to solve this problem is to throw away
these higher modes and to say that the $\{-N,...,N\}$ part generates
the regularized diffeomorphism. We can write down this as follows:
$\phi^i_N = \varphi^i_N + \epsilon\left[v^i(\varphi_N)\right]_N$
where the square brackets operation  was introduced in the 3.1.2
subsection - it projects on the $span\{e^{-i N t},...,e^{i N t}\}$
subspace. Or in terms of ($\ref{projector}$):
\begin{equation}{
\label{mode_projector} \left[ v^i\left(\varphi_N(t)\right)\right]_N
\equiv \int v^i\left(\varphi_N(t')\right)\delta_N(t'-t) dt' }
\end{equation}

So we provide the diffeomorphism of the mode-regularized theory:
\begin{equation}{
\label{reg_diff_bos} \phi^i_N = \varphi^i_N + \epsilon
\left[v^i(\varphi_N)\right]_N }
\end{equation}
\begin{equation}{
\label{reg_diff_ferm} \psi^i_N = \widetilde{\psi}^i_N + \epsilon
\left[
\partial_j v^i(\varphi_N) \widetilde{\psi}^j_N \right]_N }
\end{equation}

Which is really a diffeomorphism if $\epsilon$ is small enough
(because it is an identical diffeomorphism for $\epsilon=0$). As
long as this is nothing but the coordinate change in a
finite-dimensional integral, the answer will stay unchanged.

The measure will obviously stay invariant under such transformation.
But the inserted observables as well as the action and the
delta-function will not transform as in the continuous
(infinite-dimensional) case as we'll see later - their
transformation will give rise to some additional anomalous terms
that will explain the anomalies obtained in the Naive approach.

So we start again from the two-point correlator of the interacting
theory ($\ref{undeformed_corr}$), but written in a mode
regularization with cut-off parameter $N$:
\begin{equation}{
\label{undeformed_regular}
\langle\phi_N^a(t_1)\phi_N^b(t_2)\rangle_N={\int\prod_n d\phi_n
d\psi_n \phi_N^a(t_1)\phi_N^b(t_2)
e^{{i\over\hbar}\int\left\{{1\over
2}\omega^{(0)}_{ij}\phi_N^i\dot\phi_N^j + e_i(\phi_N)\dot\phi_N^i +
ferm.\right\}}\delta\left(\phi_N^i(\pi)\right)\over \int\prod_n
d\phi_n d\psi_n e^{{i\over\hbar}\int\left\{{1\over
2}\omega^{(0)}_{ij}\phi_N^i\dot\phi_N^j + e_i(\phi_N)\dot\phi_N^i +
ferm.\right\}}\delta\left(\phi_N^i(\pi)\right)} }
\end{equation}

And provide diffeomorphism
($\ref{reg_diff_bos}$)-($\ref{reg_diff_ferm}$). The
finite-dimensional measure $d\phi d\psi$ stays invariant while the
delta-function, the action and the observables do transform
nontrivially.

We'll pick up the first order in $\epsilon$ below and will study the
corresponding Ward identity that will definitely hold. We'll write
it in a form ``Naive Ward identity = anomalous terms'', appropriate
for the comparison with our naive Ward identity and will localize
the anomaly in such a way. However, we will not analyze all the
three non-invariance contributions mentioned above - we'll be
satisfied by finding the non-zero effect in the lowest order in
perturbations that will be the effect of only one of them .
\subsubsection{Delta-function non-invariance}
The delta-function transforms as follows:
\begin{eqnarray}
\label{reg_transform_delta}
 \delta^{(d)}\Big(\phi_N(\pi)\Big) =
\delta^{(d)}\Big(\varphi_N(\pi) +
\epsilon\left[v\left(\varphi_N(\pi)\right)\right]_N\Big)\qquad\qquad\qquad\qquad\qquad\qquad\qquad\qquad\qquad\cr
 =\delta^{(d)}\Big(\varphi_N(\pi)\Big)
+
\qquad\qquad\qquad\qquad\qquad\qquad\qquad\qquad\qquad\qquad\qquad\qquad\qquad\qquad\qquad\cr+\epsilon\sum_{i=1}^d
\delta\Big(\varphi^1_N(\pi)\Big)\delta\Big(\varphi^2_N(\pi)\Big)...\delta'\Big(\varphi^i_N(\pi)\Big)...\delta\Big(\varphi^d_N(\pi)\Big)
\Big[v^i\big(\varphi_N(\pi)\big)\Big]_N+o(\epsilon)\qquad\qquad
\end{eqnarray}
where the expansion in $\epsilon$ makes sense only inside the
integrals and presence of the delta-functions derivative means that
we should integrate by parts\footnote{In fact, the following holds:
\begin{equation}{
\label{delta_taylor} \int dx \delta\left(x + \epsilon g(x)\right)
f(x) = \int dx \delta(x) f(x) + \epsilon\int dx \delta'(x)g(x) f(x)
+ o(\epsilon)  }
\end{equation}
with appropriate restrictions on $g(x)$. In
($\ref{reg_transform_delta}$) we used this ``Taylor expansion'' for
the delta-function. }.

We substitute ($\ref{reg_transform_delta}$) into the r.h.s of
($\ref{undeformed_regular}$), integrate by parts by $\varphi^i_0$ in
the terms containing $\delta'\left(\phi^i_N(\pi)\right)$, extract
the $O(\epsilon)$ contribution and get:
$$
-\epsilon\left\langle\varphi^a_N(t_1)\varphi^b_N(t_2){\partial\over\partial\varphi^i_0}\left[v^i\left(\varphi_N(\pi)\right)\right]_N\right\rangle_{\widehat{ab}}
-
\epsilon\left\langle\left[v^a\left(\varphi_N(\pi)\right)\right]_N\varphi^b_N(t_2)\right\rangle-
\epsilon\left\langle\varphi^a_N(t_1)\left[v^b\left(\varphi_N(\pi)\right)\right]_N\right\rangle-
$$
\begin{equation}{
\label{delta_noninv}
-\epsilon\left\langle\varphi^a_N(t_1)\varphi^b_N(t_2)\left[v^k\left(\varphi_N(\pi)\right)\right]_N{i\over\hbar}\int\left\{\partial_k
e_i(\varphi_N)\dot\varphi_N^i + \partial_k
(de)_{ij}\widetilde{\psi}_N^i\widetilde{\psi}_N^j\right\}
dt\right\rangle_{\widehat{ab}} }
\end{equation}
where the first term arose from the factor
$\left[v^i\left(\varphi_N(\pi)\right)\right]_N$ in
($\ref{reg_transform_delta}$) after integration by parts, the second
and the third came from the observables and the fourth term appeared
due to the action dependence on $\varphi^i_0$. We used the notation
${\widehat{ab}}$ again to disallow contractions between
$\varphi_N^a(t_1)$ and $\varphi_N^b(t_2)$ because the appropriate
terms are canceled out due to the denominator of
($\ref{undeformed_regular}$). We can also notice that
${\partial\over\partial\varphi^i_0}\left[v^i\left(\varphi_N(\pi)\right)\right]_N=
\left[\partial_i v^i\left(\varphi_N(\pi)\right)\right]_N$. It is
convenient to rewrite ($\ref{delta_noninv}$) in a following compact
way:
\begin{equation}{
-\epsilon\left\langle D\{\phi^a_N(t_1)\phi^b_N(t_2)\} \right\rangle
}
\end{equation}
with $D$ being a differential operator:
\begin{eqnarray}
\label{delta_noninv_contr} D=\left[\partial_i
v^i\left(\varphi_N(\pi)\right)\right]_N
+\left[v^k\left(\varphi_N(\pi)\right)\right]_N{i\over\hbar}\int\left\{\partial_k
e_i(\varphi_N)\dot\varphi_N^i + \partial_k
(de)_{ij}\widetilde{\psi}_N^i\widetilde{\psi}_N^j\right\} dt +\cr
+\left[v^i\left(\varphi_N(\pi)\right)\right]_N{\partial\over\partial\varphi_0^i}\qquad\qquad
\end{eqnarray}

Notice that this really contributes because of the projector
$[..]_N$ presence: $\partial_i v^i(\varphi_N(\pi))$ and
$v^i(\varphi(\pi))$ vanish inside the correlators (due to
$\delta^{(d)}\left(\varphi(\pi)\right)$) while $\left[\partial_i
v^i(\varphi_N(\pi))\right]_N$ and $\left[v^i(\varphi(\pi))\right]_N$
don't.

The delta-function non-invariance was absent in the naive approach
at all.
\subsubsection{Action transform and its non-invariance}
After we provided the diffeomorphism
($\ref{reg_diff_bos}$)-($\ref{reg_diff_ferm}$), the change of the
action is:
$$
\Delta S_{reg} = \epsilon\int \{ \omega^{(0)}_{ij} \left[
v^i(\varphi_N) \right]_N \dot\varphi^j_N +
\left[v^k(\varphi_N)\right]_N(de(\varphi_N))_{ki}\dot\varphi^i_N +
2\omega^{(0)}_{ij}\left[\partial_k
v^i(\varphi_N)\widetilde{\psi}^k_N\right]_N\widetilde{\psi}^j_N +
$$
\begin{equation}{
\label{act_noninv}
 + \left[ v^k\right]_N\partial_k (de(\varphi_N))_{ij}\widetilde{\psi}^i_N\widetilde{\psi}^j_N + 2(de(\varphi_N))_{ij}\left[\partial_k v^i(\varphi_N) \widetilde{\psi}^k\right]_N\widetilde{\psi}^j\}dt
}
\end{equation}
Notice that $\int \left[v^i(\varphi_N(t))\right]_N\dot\varphi^j_N(t)
dt = \int \int v^i(\varphi_N(t'))\delta_N(t'-t) dt'
\dot\varphi^j_N(t) dt = \int v^i(\varphi_N(t')) \dot\varphi^j_N(t')
dt'$ and analogously $\int \left[\partial_k
v^i(\varphi_N)\widetilde{\psi}^k_N\right]_N\widetilde{\psi}^j_N dt =
\int
\partial_k v^i(\varphi_N)\widetilde{\psi}^k_N \widetilde{\psi}^j_N dt$. So we have:
$$
\Delta S_{reg} = \epsilon\int \{ \omega^{(0)}_{ij} v^i(\varphi_N)
\dot\varphi^j_N +
\left[v^k(\varphi_N)\right]_N(de(\varphi_N))_{ki}\dot\varphi^i_N +
2\omega^{(0)}_{ij}\partial_k v^i(\varphi_N)\widetilde{\psi}^k_N
\widetilde{\psi}^j_N +
$$
\begin{equation}{
\label{act_trans_noninv}
 + \left[ v^k\right]_N\partial_k (de(\varphi_N))_{ij}\widetilde{\psi}^i_N\widetilde{\psi}^j_N + 2(de(\varphi_N))_{ij}\left[\partial_k v^i(\varphi_N) \widetilde{\psi}^k\right]_N\widetilde{\psi}^j\}dt
}
\end{equation}
We see that the last expression is different from the one obtained
in the naive case (see ($\ref{act_transform}$ --
$\ref{act_trans_fermions}$)). Namely:
$$
A = {\Delta S_{naive} - \Delta S_{reg}\over \epsilon} = \int dt
\bigg\{ \left(v^k(\varphi_N) -
\left[v^k(\varphi_N)\right]_N\right)(de(\varphi_N))_{ki}\dot\varphi^i_N
+
$$
\begin{eqnarray}
\label{action_noninv_contr} + \left(v^k - \left[
v^k\right]_N\right)\partial_k
(de(\varphi_N))_{ij}\widetilde{\psi}^i_N\widetilde{\psi}^j_N+\qquad\qquad\qquad\qquad\qquad\qquad\qquad\qquad\cr
+2(de(\varphi_N))_{ij}\left(\partial_k v^i(\varphi_N)
\widetilde{\psi}^k - \left[\partial_k v^i(\varphi_N)
\widetilde{\psi}^k\right]_N\right)\widetilde{\psi}^j \bigg\}
\end{eqnarray}

This difference will be referred to as the action non-invariance.
Notice that it naively vanishes in the $N\to\infty$ limit. However
it still can contribute if one takes off the regularization
carefully - at the end of the computations.

\subsubsection{Observables non-invariance}
The observables transform as follows:
\begin{equation}{
\label{reg_obs_transform} \phi^a_N(t_1)\phi^b_N(t_2) \to
\varphi^a_N(t_1)\varphi^b_N(t_2) +
\epsilon\varphi^a_N(t_1)\left[v^b\left(\varphi_N(t_2)\right)\right]_N
+
\epsilon\left[v^a\left(\varphi_N(t_1)\right)\right]_N\varphi^b_N(t_2)
}
\end{equation}
Non-invariance generated by this transformation, namely the
difference from the naive setup case is:
\begin{eqnarray}
\label{obs_noninv} O={naive - proper\over\epsilon}
=\varphi^a_N(t_1)\left(v^b\left(\varphi_N(t_2)\right)-\left[v^b\left(\varphi_N(t_2)\right)\right]_N\right)
+\cr
+\left(v^a\left(\varphi_N(t_1)\right)-\left[v^a\left(\varphi_N(t_1)\right)\right]_N\right)\varphi^b_N(t_2)
\end{eqnarray}
Again it naively vanishes in the $N\to \infty$ limit.

Our following step will be to gather the $O(\epsilon)$ terms
altogether and to obtain the Ward identities. As long as in the
proper approach we make diffeomorphism of the finite-dimensional
(i.e. well-defined) integral, these identities will really hold and
will show why the naive approach failed.

\subsubsection{Ward identity}
Let us gather together all the terms above and write the
$O(\epsilon)$ contribution. It should vanish and certainly will.
We'll write it in the form ``naive Ward identity (N.W.I.)'' $=$
``proper non-invariant terms'':
\begin{equation}{
\label{fair_ward_id} N.W.I.=\left\langle
D\left\{\varphi_N^a(t_1)\varphi_N^b(t_2)\right\}\right\rangle_N^{\widehat{ab}}
+ \left\langle A \varphi_N^a(t_1)\varphi_N^b(t_2)
\right\rangle_N^{\widehat{ab}} + \left\langle O \right\rangle_N }
\end{equation}
where ``N.W.I.'' equals to the l.h.s of ($\ref{naive_ward_id}$).
Notice that $D$ is a differential operator and we have:
\begin{eqnarray}
\label{delta_noninv_contr_detailed} \left\langle
D\left\{\varphi_N^a(t_1)\varphi_N^b(t_2)\right\}\right\rangle_{\widehat{ab}}=\left\langle\varphi^a_N(t_1)\varphi^b_N(t_2)\left[\partial_i
v^i\left(\varphi_N(\pi)\right)\right]_N\right\rangle_{\widehat{ab}}
+\cr
+\left\langle\left[v^a\left(\varphi_N(\pi)\right)\right]_N\varphi^b_N(t_2)\right
\rangle+\left\langle\varphi^a_N(t_1)\left[v^b\left(\varphi_N(\pi)\right)\right]_N\right\rangle+\cr
+\left\langle\varphi^a_N(t_1)\varphi^b_N(t_2)\left[v^k\left(\varphi_N(\pi)\right)\right]_N{i\over\hbar}\int\left\{\partial_k
e_i(\varphi_N)\dot\varphi_N^i + \partial_k
(de)_{ij}\widetilde{\psi}_N^i\widetilde{\psi}_N^j\right\}
dt\right\rangle_{\widehat{ab}}
\end{eqnarray}
So we have three possible anomalous terms - $D$ describes
delta-function non-invariance, $A$ describes action non-invariance
and $O$ describes observables non-invariance.
\subsubsection{The zeroth order identity}
Similar to the naive setup, we pick up the zeroth order in
$e_i(\varphi)$ from ($\ref{fair_ward_id}$). Let us analyze different
contributions:

1) $O$ from ($\ref{obs_noninv}$) doesn't contribute as
$$
\langle\varphi_N^a(t_1) \left[v^b(\varphi_N(t_2)) \right]_N\rangle_N
= \int \langle\varphi^a(t_1)v^b(\varphi_N(t))\rangle_N
\delta_N(t-t_2) dt
$$
and in the zeroth order in $e_i(\varphi)$ we have
\begin{equation}{
\label{self_vanishing}
^0\left\langle\varphi^a(t_1)v^b(\varphi_N(t))\right\rangle_N=0 }
\end{equation}
as long as $v^b(\varphi)$ starts from the quadratic terms and we
have $^0\langle\varphi_N^i(t)\varphi_N^j(t)\rangle_N=0$ in our
prescription. Moreover, in our case $O$ won't contribute in all
orders in $e_i(\varphi)$ as long as
$\left[v^a\left(\varphi_N(t)\right) \right]_N -
v^a\left(\varphi_N(t)\right)$ contains only higher Fourier modes in
$t$ - modes with absolute value of mode number greater than $N$. But
this means that every Fourier mode with finite number vanishes in
the $N\to\infty$ limit and if the answer exists at all - it should
be zero. In general we could consider such observables that this
argument wouldn't be true (e.g. integrated observables), but in our
case $O$ doesn't contribute.

2) $D$ from ($\ref{delta_noninv_contr_detailed}$) contributes but
only the first term is relevant. The second and the third terms
vanish in the zeroth order for the same reasons as
($\ref{self_vanishing}$). The last term in
($\ref{delta_noninv_contr_detailed}$) is of the higher order in
$e_i(\varphi)$ and we neglect it.

3) $A$ from ($\ref{action_noninv_contr}$) is of the higher order in
$e_i(\varphi)$ and thus is not relevant.

4) The zeroth order (in $e_i(\varphi)$) part of the l.h.s. of
($\ref{fair_ward_id}$), i.e. the zeroth order part of the naive Ward
identity ($\ref{naive_ward_id}$) has already been calculated in the
4.1.2 subsection and is given by ($\ref{ward_id_breakdown}$).

So after all we are left with the following terms:
\begin{equation}{
\label{zeroth_identity} -\hbar^2\theta_{(0)}^{a i_1}\theta_{(0)}^{b
i_2}
\partial_{i_1}\partial_{i_2}\partial_i v^i(0) A^{(2)}(t_1,t_2)=\
^0\left\langle\varphi^a_N(t_1)\varphi^b_N(t_2) \left[\partial_i
v^i\left(\varphi_N(\pi)\right)\right]_N\right\rangle^{\widehat{ab}}
}
\end{equation}

But one can notice that $\left[\partial_i
v^i(\varphi(\pi))\right]_N$ is nothing but the anomalous vertex
($\ref{anom_vert_big}$) where one has replaced $e_i(\varphi)$ by
$\omega^{(0)}_{ki}v^k$. Thus we can use the result
($\ref{twolegs_answer}$) for the r.h.s. of ($\ref{zeroth_identity}$)
and find that ($\ref{zeroth_identity}$) really holds - the anomaly
of the naive Ward identity is canceled out by the term
$\left\langle\varphi^a_N(t_1)\varphi^b_N(t_2) \left[\partial_i
v^i\left(\varphi_N(\pi)\right)\right]_N\right\rangle_0^{\widehat{ab}}$
which describes delta-function non-invariance.

This shows that in the lowest order in perturbations the Ward
identity holds. Although we are not analyzing the higher orders,
it's clear that it holds there too, just because the proper
prescription deals with finite-dimensional space of fields.

\section{Discussion and conclusions}
The current paper was devoted to the study of the general phase
space covariance of quantum mechanics. We have been working in the
path integral formalism with a special model theory
($\ref{naive_product}$). During this study we have seen that the
naive bosonic path integral ($\ref{naive_product}$) (in the phase
space) is an ill-defined object as it gives rise to an extremely
divergent theory. The solution of the problem was to
``supersymmetrize'' it by adding anticommuting ghost fields
transforming in a proper way, such that the modified theory became
finite and well-defined. At the end of the day we found that the
quantum answer was not invariant under classical diffeomorphisms -
it behaved anomalously. And the source of the anomaly was the whole
path integral becoming non-invariant after regularization.

In the first half of the paper the super-improvement was discussed.
The framework was formulated and its self-consistency was proven.
The main idea of the method was to replace a naive bosonic measure
by a super-modified one:
\begin{equation}{
\label{measure_repl} \mathcal{D} \phi \to
\mathcal{D}\phi\mathcal{D}\psi\ e^{{i\over\hbar}\int
\omega_{ij}(\phi)\psi^i\psi^j dt} }
\end{equation}
Then some illustrative examples as well as an important discussion
of the anomalous vertex were performed.

From the ``Examples'' chapter we have studied that our theory has a
peculiar issue connected with the specific behavior of correlators.
Correlators of observables in the phase space contain jumps in their
time dependences. The value of the given correlation function at the
point of this jump is not fixed in general and depends on the
regularization (for example it depends on the form of UV cut-off in
our case). When providing perturbative analysis one often has to
compute the integrals of the form $\int f(t)\delta(t-t_0) dt$, where
the function $f(t)$ contains jump at the point $t=t_0$ -- we face
such troubles when considering $A^{(n)}(t_1,...,t_n)$. Such
integrals need regularization badly and their values depend on it.

Then the problem of general covariance was discussed. We were
interested in a diffeomorphism action in our theory. The naive
approach to the subject was to make a classical diffeomorphism
($\ref{diff_boson_s1}$, $\ref{diff_fermion_s1}$) in our super-space
first, not worrying about regularization at all. Such way of
thinking gave rise to the anomalous Ward identity
($\ref{naive_ward_id}$). In the lowest order in perturbations it was
found (see ($\ref{naive_zeroth}$)) that the source of the anomaly in
such approach was the anomalous vertex had been obtained before. We
have seen that, roughly speaking, the ambiguity appearing when
integrating jump with the delta-function generated the anomaly in
the lowest order.

However then we described a proper prescription, which demonstrated
the real origin of non-invariance. The proper prescription said,
that one had to introduce the regularization first and then work in
terms of the regularized finite-dimensional theory. The notion of
diffeomorphism had to be modified as long as the classical
diffeomorphisms didn't respect the mode-regularization - they
spoiled the cut-off, generating modes with mode number above the
cut-off parameter $N$. That's why we had to introduce the
``regularized'' diffeomorphisms (\ref{reg_diff_bos},
$\ref{reg_diff_ferm}$) - finite-dimensional diffeomorphisms of the
regularized theory imitating classical ones in the $N\to\infty$
limit. However all the components of the path integral construction
became non-invariant after such modification, or more precisely,
they transformed in a non-classical, non-covariant way under these
regularized diffeomorphisms (for the fixed finite cut-off parameter
$N$). These non-covariances are treated as the real origin of the
anomaly in a mode-regularization prescription.

We have computed separately all the potentially anomalous terms -
the delta-function anomaly $D$ (see ($\ref{delta_noninv_contr}$)),
the action anomaly $A$ (see ($\ref{action_noninv_contr}$)) and the
observables anomaly $O$ (see (\ref{obs_noninv})). All these three
terms could in principle contribute if we considered higher orders
of perturbation theory and arbitrary observables. However we were
satisfied by locating the lowest non-trivial anomalous contribution
- it came from the delta-anomaly $D$ in our case. Notice that in
general one could replace the delta-function zero-mode regulator
(the evaluation observable) by a non-zero Hamiltonian. In such case
the delta-function non-invariance would be replaced by the
Hamiltonian non-invariance (which would be almost the same as the
action non-invariance).

In general it would be interesting and important to study the higher
orders (where $A$ and for special observables also $O$ starts to
contribute) and to explore how the classical symmetries are deformed
at the quantum level and give rise to the Kontsevich's
$L_{\infty}$-morphism. But we leave this beyond the current paper as
a topic of an upcoming research.

One could ask an interesting question, whether the claimed anomaly
was an artifact of the regularization (which by itself was
non-invariant under diffeomorphisms), or it was a fundamental
property of the theory. We know that there is no quantum answer for
the problem ($\ref{naive_product}$) that could be invariant under
classical diffeomorphisms. Thus we could conclude that the anomaly
was a fundamental property of the theory. And what we have done is
just the analysis of this anomaly in a mode-regularization
prescription. However it would be much better to study the subject
in another framework with some alternative regularization as well in
order to understand if anything depends on it. This alternative
regularization is time-slicing (or equivalently discretization, or
lattice regularization). In such case one replaces continuous time
by the set of $N$ points, continuous fields $\phi^i(t),\ \psi^i(t)$
-- by the set of their values at these points: $\phi^i_k,\ \psi^i_k$
and considers the following object:
\begin{equation}{
\label{discreete} \int \prod_{k,i} d\phi^i_k d\psi^i_k
\prod_i\delta(\phi^i_N-x^i) \exp{{i\over\hbar}\sum_k \left\{
\alpha_i(\phi_k)(\phi^i_{k+1}-\phi^i_k) +
\omega_{ij}(\phi_k)\psi^i_k\psi^j_k \right\}}}
\end{equation}
where measure, delta-function and
$\omega_{ij}(\phi_k)\psi^i_k\psi^j_k$ term are invariant under
classical diffeomorphisms! The only term that behaves in a
non-classical way is $\alpha_i(\phi_k)(\phi^i_{k+1}-\phi^i_k)$.
However despite of this seeming simplification, the time-slicing
approach needs some extra analysis and additional assumptions as in
our special case the proper continuous limit $N\to\infty$ doesn't
exist even for the \emph{quadratic} theory! This is connected with
the delta-function making an illegal operation - fixing coordinates
and momentums at the same point. So one needs to think over this
subtle issue carefully and thus we leave this for an upcoming
research as well.
\appendix
\section{Bosonic divergences\\}
First we need to find a propagator:
\begin{equation}{
^0\label{a_prop_def} \langle\phi^i(t_1)\phi^j(t_2)\rangle = {\int
\prod_{n,i}d\phi^i_n \prod_{i=1}^d\delta(\sum_n \phi^i_n (-1)^n-x^i)
\phi ^i (t_1)\phi^j(t_2) e^{-{1\over\hbar}\sum_n
\pi\omega^{(0)}_{ij}\phi^i_{-n}n\phi^j_n} \over \int
\prod_{n,i}d\phi^i_n \prod_{i=1}^d\delta(\sum_n \phi^i_n (-1)^n-x^i)
e^{-{1\over\hbar}\sum_n \pi\omega^{(0)}_{ij}\phi^i_{-n}n\phi^j_n}} }
\end{equation}
\begin{equation}{
^0\label{a_prop_val} \langle\phi^i(t_1)\phi^j(t_2)\rangle =x^i x^j +
i\hbar\theta^{ij}_{(0)} G(t_1,t_2) }
\end{equation}
Where $\theta_{(0)}=\left(\omega^{(0)}\right)^{-1}$,
$G(t_1,t_2)=\sum_{n,m} G_{n,m} e^{i n t_1 + i m t_2}$ and:
$$
G_{n,m}={1\over 2\pi i}\left\{ {\delta_{n+m}\over n} -
{\delta_m\over n}(-1)^n + {\delta_n\over m}(-1)^m\right\}
$$
\begin{equation}{
\label{a_propagator} G_{0,0}=0 }
\end{equation}
Notice that we think of it as of an infinite-dimensional matrix, inverse to $\pi\omega_{ij}^{(0)}n\delta_{n+m}$ - again we don't care about correctness yet.\\
From ($\ref{alpha_form}$) we see that the only interaction in our theory is $e_i(\phi)\dot\phi^i$ which is in fact a series of interactions: $e_i(\phi)\dot\phi^i={1\over 2}\partial_{a_1}\partial_{a_2}e_i(0)\phi^{a_1}\phi^{a_2}\dot\phi^i + {1\over 3!}\partial_{a_1}\partial_{a_2}\partial_{a_3}e_i(0)\phi^{a_1}\phi^{a_2}\phi^{a_3}\dot\phi^i+...$ (we can neglect linear term due to the periodic boundary conditions in time and get rid of the quadratic term by relating it to the free part of the action).\\
It makes sense to introduce diagram rules for the further considerations.\\
\textbf{Diagram rules in the coordinate space}\\
Let us denote a free propagator by the line:
\begin{equation}{
^0\label{a_feyn_prop_coord} \langle \phi^i(t_1)\phi^j(t_2) \rangle =
\epsfig{file=prop1.eps} }
\end{equation}
and introduce a convenient notation for the interaction - a vertex
with an arrow on the leg containing the time derivative (as in
[[article]]):
\begin{equation}{
\label{a_feyn_vertex_coord} {1\over
k!}\partial_{a_1}...\partial_{a_k}e_i(0)\phi^{a_1}(t)...\phi^{a_k}(t)\dot\phi^i(t)
= \epsfig{file=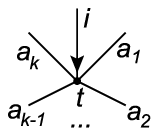} }
\end{equation}
\textbf{Diagram rules in the momentum space}\\
Again denote a free propagator by the line:
\begin{equation}{
^0\label{a_feyn_prop_mom} \langle \phi^i_n\phi^j_m \rangle =
\epsfig{file=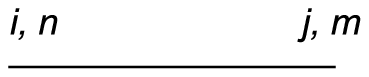} }
\end{equation}
and interaction - by the vertex with an arrow on the leg containing
the time derivative:
\begin{equation}{
\label{a_feyn_vertex_mom} {2\pi\over
k!}\partial_{a_1}...\partial_{a_k}e_i(0)\delta_{n_1+n_2+...+n_k+s}\phi^{a_1}_{n_1}...\phi^{a_k}_{n_k}s\phi^i_{s}
= \epsfig{file=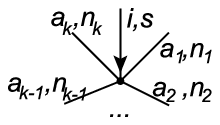} }
\end{equation}
Now let us try to compute an interacting propagator up to the second
order, or at least observe, how these computations could be
provided. For simplicity assume that
$e_i(\phi)=a_{i,j,k}\phi^j\phi^k$ contains only quadratic terms.
Then an interacting propagator is given by the following series of
graphs:
\begin{eqnarray}
\label{a_feyn_inter} ^0\langle\phi^i(t_1)\phi^j(t_2)e^{{i\over\hbar}
\int e}\rangle
\qquad\qquad\qquad\qquad\qquad\qquad\qquad\qquad\qquad\qquad\qquad\qquad\qquad\qquad\cr
\begin{matrix}=\cr\ \cr\end{matrix}\includegraphics[scale=1]{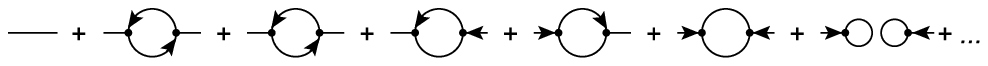}\qquad
\end{eqnarray}
As we see, they contain several types of loops. These loops bring
some divergences and ambiguities. For example the loop:
$$
\epsfig{file=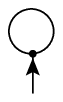}
$$
corresponds in a momentum space to $\sum_{n,m}s\phi^i_s\delta_{n+m+s}G_{n,m}={s\phi^i_s\over 2\pi i}\delta_s\sum_n{1\over n}$ which is a logarithmic divergent expression. But here we can notice some effect - in the case of an odd-dimensional space-time (which is our case) logarithmic divergences are not real divergences - they don't need any counterterms to cancel them, all we need to do is to provide some kind of ultroviolet regularization - an UV cut-off in our case - and the logarithmic divergency won't take place any more. Instead of it a logarithmic ambiguity appears - the answer depends on the form of the UV cut-off. This shows us that at least we have to replace a formal infinite sum in ($\ref{fourier1}$) by the finite one with some limiting UV parameter $N$.\\
The loop of the type:
$$
\epsfig{file=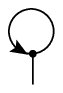}
$$
As well as the loops:
$$
\epsfig{file=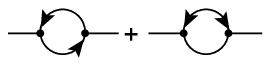}
$$
are not logarithmic but linear divergent! Indeed, every propagator
is proportional to the inverse power of the momentum and every arrow
on the leg (indicating derivative) is ``proportional'' to the first
power of the momentum. Thus as soon as the loop contains equal
numbers of propagators and arrows, it can produce a linear
divergence. Indeed, the first loop is proportional to
$\sum_{n,m}\delta_{n+m+s}n G_{nm}={\delta_s\over 2\pi i}\sum_n 1 -
{(-1)^s\over 2\pi i}$ which is a linear divergence. The same can be
obtained for the second one. These are real divergences that should
be cancelled by some manually added counterterms at this point of
consideration.

The situation becomes really weird if we notice that there exists an
infinite series of wheel-like diagrams all being linearly divegrent:
$$
\epsfig{file=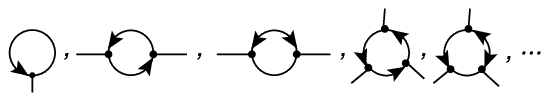}
$$
such series often indicates that the theory is non-renormalizable as
long as every set of diagrams of the series with the fixed number of
external legs needs a new type of counterterm to be introduced. It
looks natural and strange at once. People know - interactions
containing derivatives often happen to be non-renormalizable, but
they are also used to the fact that one-dimensional theories are
finite. So what we have to do?

Again we see that we need to provide an ultraviolet cut-off to make
a sence of all this. And more - we need to add counterterms. The
good thing is that all the necessary counterterms are encoded in
($\ref{covar_measure}$) which leads us to the super-improvement
idea.
\section{Cancelation of divergences}
\textbf{Lemma 1:} Every non-ghost loop that contains less number of
internal arrows (indicating time derivatives) than internal
propagators is finite. For example:
$$
\epsfig{file=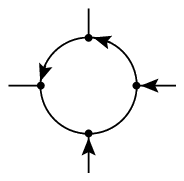}
$$
\textbf{Sketch of proof:} Notice that an ultraviolet behavior
doesn't depend on the infrared one and thus we can think for a
moment that $G_{nm}={1\over 2\pi i}{\delta_{n+m}\over n}$, i.e. that
the bosonic propagator conserves momentum (one can check if needed
that this really doesn't affect the divergences). After that we can
provide the following argument: suppose the momentum $p$ circulates
over the loop; when $p\to\infty$ every propagator carries the factor
${1\over p}$ and every internal arrow gives rise to the factor $p$;
then as soon as we have more propagators than arrows, total loop is
proportional at least to the first inverse power of $p$; thus it
converges in a sense of mode regularization due to the Observation 4
- because in our theory expressions like $\sum{1\over p}$ are
finite.
\\

Now consider a ghost loop with $n$ vertexes:
$$
\epsfig{file=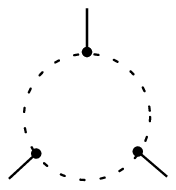}
$$
\textbf{Lemma 2:} this diagram cancels the divergent part of the sum
of non-ghost loops with $n$ vertexes and with equal numbers of
internal propagators and internal arrows:
$$
\epsfig{file=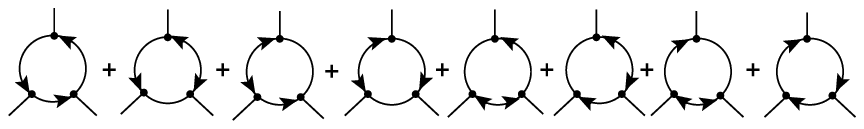}
$$
\textbf{Proof: }Let us introduce a convenient notation for such
diagrams. Go around the diagram in a clockwise direction, numerate
vertexes and for the $k$'th vertex say that it is of the type "c" if
the appropriate arrow is in the clockwise direction to the vertex
and we say that it is of the type "a" if the appropriate arrow is in
the anticlockwise direction to the vertex. We also associate indexes
$i_k$ and $j_k$ to each vertex ($i$ - to the internal leg without
arrow and $j$ - to the internal leg with arrow). Denote such a loop
as follows: $Loop\left[^{i_1 j_1}_{d_1},^{i_2 j_2}_{d_2},...,^{i_n
j_n}_{d_n}\right]$ where $d_k$ indicates the type of the vertex and
is either "c" or "a". We'll also associate an analytical expression
to the $Loop[\ ]$ in a natural way:
\begin{equation}{
\label{b_loop_def} Loop\left[^{i_1 j_1}_{c },^{i_2 j_2}_{c
},...,^{i_n j_n}_{c }\right]=\
^0\langle\dot\phi^{j_1}(t_1)\phi^{i_2}(t_2)\rangle \
^0\langle\dot\phi^{j_2}(t_2)\phi^{i_3}(t_3)\rangle...^0\langle\dot\phi^{j_n}(t_n)\phi^{i_1}(t_1)\rangle
}
\end{equation}
which is nothing but the product of internal propagators along the
loop. In terms of $Loop[\ ]$ we can write the total contribution of
our loops as follows (we don't mind about external legs explicitly):
\begin{equation}{
\label{b_loops_contr}
^0\left\langle...\left({i\over\hbar}\right)^n\int
dt_1...dt_n\sum_{d_k\in\{c,\
a\}}\partial_{i_1}e_{j_1}\left(\phi(t_1)\right)...
\partial_{i_n}e_{j_n}\left(\phi(t_n)\right)Loop\left[^{i_1
j_1}_{d_1},^{i_2 j_2}_{d_2},...,^{i_n
j_n}_{d_n}\right]...\right\rangle }
\end{equation}
Notice that:
$$
\int dt_1 \partial_{i_1}e_{j_1} Loop\left[^{i_1 j_1}_{a},^{i_2
j_2}_{c},...,^{i_n j_n}_{c}\right]
$$
$$
=\int dt_1 \partial_{i_1}e_{j_1} \
^0\langle\phi^{i_1}(t_1)\phi^{i_2}(t_2)\rangle\
^0\langle\dot\phi^{j_2}(t_2)\phi^{i_3}(t_3)\rangle...^0\langle\dot\phi^{j_n}(t_n)\dot\phi^{j_1}(t_1)\rangle=({\rm
intagrate\ by\ parts})
$$
$$
=-\int dt_1 \partial_{i_1}e_{j_1} \
^0\langle\dot\phi^{i_1}(t_1)\phi^{i_2}(t_2)\rangle\
^0\langle\dot\phi^{j_2}(t_2)\phi^{i_3}(t_3)\rangle...^0\langle\dot\phi^{j_n}(t_n)\phi^{j_1}(t_1)\rangle-
$$
$$
-\int dt_1 \dot\phi^k(t_1)\partial_p\partial_{i_1}e_{j_1} \
^0\langle\phi^{i_1}(t_1)\phi^{i_2}(t_2)\rangle\
^0\langle\dot\phi^{j_2}(t_2)\phi^{i_3}(t_3)\rangle...^0\langle\dot\phi^{j_n}(t_n)\phi^{j_1}(t_1)\rangle=(i_1\leftrightarrow
j_1)
$$
\begin{equation}{
\label{b_int_by_parts} =-\int dt_1 \partial_{j_1}e_{i_1}
Loop\left[^{i_1 j_1}_{c},^{i_2 j_2}_{c},...,^{i_n j_n}_{c}\right] +
{\rm regular\ part}
 }
\end{equation}
Now we can see that:
\begin{eqnarray}
\label{b_derham_appears} &&\int dt_1 \partial_{i_1}e_{j_1} \left(
Loop\left[^{i_1 j_1}_{a},^{i_2 j_2}_{c},...,^{i_n j_n}_{c}\right] +
Loop\left[^{i_1 j_1}_{c},^{i_2 j_2}_{c},...,^{i_n
j_n}_{c}\right]\right)\cr &&=\int dt_1
\left(\partial_{i_1}e_{j_1}-\partial_{j_1}e_{i_1}\right)
Loop\left[^{i_1 j_1}_{c},^{i_2 j_2}_{c},...,^{i_n j_n}_{c}\right] +
{\rm regular\ part}\cr &&=\int dt_1 de_{i_1 j_1} Loop\left[^{i_1
j_1}_{c},^{i_2 j_2}_{c},...,^{i_n j_n}_{c}\right] + {\rm regular\
part}
\end{eqnarray}
Analogously one can show that:
\begin{equation}{
\label{b_loops_contr_modif} (\ref{b_loops_contr})= \
^0\left\langle...\left({i\over\hbar}\right)^n\int dt_1...dt_n
(de)_{i_1 j_1}\left(\phi(t_1)\right)... (de)_{i_n
j_n}\left(\phi(t_n)\right)Loop\left[^{i_1 j_1}_{c},...,^{i_n
j_n}_{c}\right]...\right\rangle}+ {\rm reg}
\end{equation}
Naively one can see that $Loop\left[^{i_1 j_1}_{c},...,^{i_n
j_n}_{c}\right]\sim\delta(0) + reg$. Let us calculate it in a
mode-regularization framework. Consider a loop in the momentum
space:
$$
\epsfig{file=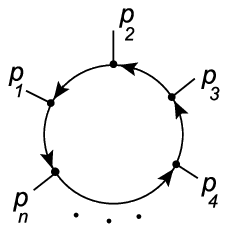}
$$
To calculate its divergent part we can again put $G_{n,m}={1\over
2\pi i}{\delta_{n+m}\over n},\ G_{0,0}=0$ and also vanish external
momentums $p_k=0$. Then we can put $^0\langle(\dot\phi^i)_n
\phi^j_m\rangle=i\hbar\theta_{(0)}^{ij}{\delta_{n+m}\over 2\pi}$ and
finally get an expression for the loop:
\begin{equation}{
\label{b_loop_expr} \left({i\hbar\over
2\pi}\right)^n\theta_{(0)}^{j_1 i_2}\theta_{(0)}^{j_2
i_3}\ldots\theta_{(0)}^{j_n
i_1}\sum_{\{q\},\{m\}}\delta_{q_1+m_2}\delta_{q_2+m_3}\ldots\delta_{q_n+m_1}
\delta_{q_1+m_1}\delta_{q_2+m_2}\ldots\delta_{q_n+m_n} }
\end{equation}
Here $\delta_{q_1+m_1},\ \delta_{q_2+m_2},\ldots,\delta_{q_n+m_n}$
come from the vertexes. Providing trivial summation one gets:
\begin{equation}{
\label{b_bos_loop_val} (\ref{b_loop_expr})= \left({i\hbar\over
2\pi}\right)^n\theta_{(0)}^{j_1 i_2}\theta_{(0)}^{j_2
i_3}\ldots\theta_{(0)}^{j_n i_1}\sum_{^{q=-N..N}_{q\ne 0}} 1=
\left({i\hbar\over 2\pi}\right)^n\theta_{(0)}^{j_1
i_2}\theta_{(0)}^{j_2 i_3}\ldots\theta_{(0)}^{j_n i_1} 2N }
\end{equation}
that is a linear divergent expression. Now let us consider a ghost
loop. It's contribution:
\begin{eqnarray}
\label{b_ghost_loop}
^0\bigg\langle\ldots\left({i\over\hbar}\right)^n\int dt_1\ldots dt_n
de_{i_1 j_1}\left(\phi(t_1)\right)\ldots de_{i_n
j_n}\left(\phi(t_n)\right)(-1)^{2n-1}2^n\times\cr
\times\left[^0\langle\psi^{j_1}\psi^{i_2}\rangle^0\langle\psi^{j_2}\psi^{i_3}\rangle\ldots^0\langle\psi^{j_n}\psi^{i_1}\rangle\right]\ldots\bigg\rangle
\end{eqnarray}
where factor $(-1)^{2n-1}=-1$ reflects the parity of permutation we
have to perform to transform
$\psi^{i_1}\psi^{j_1}\psi^{i_2}\psi^{j_2}...\psi^{i_n}\psi^{j_n}$
into
$\psi^{j_1}\psi^{i_2}\psi^{j_2}\psi^{i_3}...\psi^{j_n}\psi^{i_1}$
and factor $2^n$ reflects an additional symmetry of the diagram: we
can change $\psi^{i_k}\leftrightarrow\psi^{j_k}$ as they contribute
into the vertex in a similar way. Divergent part of
($\ref{b_ghost_loop}$) is proportional to $\delta(0)$, now we want
to show that it is exactly the divergence of
($\ref{b_loops_contr}$). To do that we calculate the ghost loop in
the momentum space (using ghost propagator):
\begin{equation}{
\label{b_ghost_loop_mom} -2^n \left({i\hbar\over 4\pi}\right)^n
\theta_{(0)}^{j_1 i_2}\theta_{(0)}^{j_2 i_3}\ldots\theta_{(0)}^{j_n
i_1}\sum_{\{q\},\{m\}}\delta_{q_1+m_2}\delta_{q_2+m_3}\ldots\delta_{q_n+m_1}
\delta_{q_1+m_1}\delta_{q_2+m_2}\ldots\delta_{q_n+m_n} }
\end{equation}
Here again $\delta_{q_1+m_1},\
\delta_{q_2+m_2},\ldots,\delta_{q_n+m_n}$ come from the vertexes.
After trivial summation:
\begin{equation}{
\label{b_ghost_loop_val} (\ref{b_ghost_loop_mom})=
-\left({i\hbar\over 2\pi}\right)^n\theta_{(0)}^{j_1
i_2}\theta_{(0)}^{j_2 i_3}\ldots\theta_{(0)}^{j_n i_1}\sum_{q=-N..N}
1= -\left({i\hbar\over 2\pi}\right)^n\theta_{(0)}^{j_1
i_2}\theta_{(0)}^{j_2 i_3}\ldots\theta_{(0)}^{j_n i_1} (2N+1)}
\end{equation}
Now we see that ($\ref{b_ghost_loop_val}$) and
($\ref{b_bos_loop_val}$) differ by sign and also by the finite
contribution of ghost's zero mode. Then when ($\ref{b_loops_contr}$)
and ($\ref{b_ghost_loop}$) are taken together, divergent parts do
cancel each other and finally we get
the statement of the Lemma 2.\\
\textbf{Observation 5:} Now when we have Lemma 1 and Lemma 2 in our
disposal, we automatically have the fact that any correlation
function in momentum space
$\langle\phi^{i_1}_{m_1}...\phi^{i_k}_{m_k}\rangle$ is finite.
Indeed, application of Lemma 2 gives that we can get rid of ghosts
at all together with non-ghost loops with equal numbers of
propagators and internal arrows (note that number of internal arrows
is less or equal to the number of propagators). After all Lemma
1 guarantees that all that is left is finite.\\

\end{document}